\documentclass[11pt]{article}
\usepackage{scicite}

\usepackage{times}

\usepackage{amsmath,amssymb,amsfonts}

\usepackage{graphicx}
\usepackage[colorlinks,
	linkcolor=red,
	citecolor=blue,
	urlcolor=red]{hyperref}
\graphicspath{{../Figures/}}

\usepackage{cleveref}

\newcommand{\modif}[1]{{\color{black}#1}}

\newenvironment{sciabstract}{%
\begin{quote} \bf}
{\end{quote}}
\title{Bell correlations between light and vibration}

\author{\\
Santiago Tarrago Velez$^{1}$, Vivishek Sudhir$^{2,3}$, \\Nicolas Sangouard$^{4\ast}$, Christophe Galland$^{1\ast}$\\
\normalsize{$^{1}$Institue of Physics, Ecole Polytechnique F\'{e}d\'{e}rale de Lausanne, CH-1015 Lausanne,}\\
\normalsize{Switzerland}\\
\normalsize{$^{2}$LIGO Laboratory, Massachusetts Institute of Technology, Cambridge,}\\
\normalsize{ MA 02139, USA}\\
\normalsize{$^{3}$Department of Mechanical Engineering, Massachusetts Institute of Technology,}\\
\normalsize{ Cambridge, MA 02139, USA}\\
\normalsize{$^{4}$Departement Physik, Universit\"{a}t Basel, Klingelbergstrasse 82,}\\
\normalsize{CH-4056 Basel, Switzerland} \\
\normalsize{$^{5}$Universit\'{e} Paris-Saclay, CEA, CNRS, Institut de physique th\'{e}orique,}\\
\normalsize{91191, Gif-sur-Yvette, France}\\
\normalsize{$^\ast$To whom correspondence should be addressed:  nicolas.sangouard@unibas.ch; chris.galland@epfl.ch}\\
}

\date{}

\begin{document} 
\baselineskip14pt
\maketitle 

\begin{sciabstract}


Time-resolved Raman spectroscopy techniques offer various ways to study the dynamics of molecular vibrations in liquids or gases and optical phonons in crystals.
While these techniques give access to the coherence time of the vibrational modes, they are not able to reveal the fragile quantum correlations that are spontaneously created between light and vibration during the Raman interaction. 
Here, we present a new scheme leveraging universal properties of spontaneous Raman scattering to demonstrate for the first time Bell correlations between light and a collective molecular vibration. 
We measure the decay of these hybrid photon-phonon Bell correlations with sub-picosecond time-resolution and find that they survive over several hundred oscillations at ambient conditions.
Our method offers a universal approach to generate entanglement between light and molecular vibrations. 
Moreover, our results pave the way for the study of quantum correlations in more complex solid-state and molecular systems in their natural state.
\end{sciabstract}

%
%
%
%

\section*{Introduction}
In the hierarchy of non-classical states, the Bell correlated states represent an extreme. 
When two parties share such a state, information can be encoded exclusively in the quantum correlations of the random outcomes
of measurements between them \cite{bell1964einstein,clauser1969proposed}. 
The strength of such correlations is quantified by Bell inequalities, whose violation demarcates Bell correlated states
from less entangled ones \cite{Brunner14}. 

Experimental realizations of Bell correlated states --- whether between polarization states of light \cite{freedman1972,AspRog81}, 
individual atomic systems\cite{chou2007,hofmann2012,blatt_entangled_2008}, 
in atomic ensembles \cite{matsukevich2005,kuzm13,engelsen17}, superconducting ciruits \cite{ansmann2009violation,carlo10}, 
or solid-state spins \cite{morello16,hensen2015loophole} --- call for isolated systems that strongly interact with a 
well-characterized probe. 
Even mesoscopic acoustic resonators have been engineered to exhibit Bell correlations \cite{marinkovic2018} thanks to long coherence times (achieved at milli-Kelvin temperatures) and strong interaction with light (by integration with an optical micro-cavity).

Intriguingly, recent experiments have shown that high-frequency vibrations of bulk crystals \cite{lee2011science,jorio2015,kasperczyk2015,england2016,hou2016,anderson2018} or molecular ensembles \cite{bustard2015,kasperczyk2016,saraiva2017} 
can mediate non-classical intensity correlations between inelastically scattered photons under ambient conditions (i.e. at room temperature and atmospheric pressure).   
In the pioneering work of Lee et al. \cite{lee2011science}, 
{two phonon modes in spatially separated bulk diamonds had been entangled with each other by performing coincidence measurements and post-selection on the Raman-scattered photons.
Recently, leveraging a new two-tone pump-probe method \cite{anderson2018}, it became possible to follow the birth and death of an individual quantum of vibrational energy (i.e. Fock state) excited in a single spatio-temporal  mode of vibration in a bulk crystal \cite{tarrago19}.}

Remarkably, these experiments did not necessitate specially engineered subjects; they reveal fundamental quantum properties of naturally occurring materials.
Taken together, these developments raise new questions: 
{Are the correlations spontaneously created between light and vibration during Raman scattering strong enough to violate Bell inequalities? 
How is the vibrational coherence time reflected in the dynamics of the hybrid light-vibration quantum correlations?}

In this Letter, we demonstrate for the first time Bell correlations arising from the Raman interaction between light and mechanical vibration at ambient conditions, and use them to resolve the decoherence of the  vibrational mode mediating these correlations.
While this proof-of-principle experiment is realized on a vibrational mode in a bulk diamond crystal, the effect that is revealed should be universally observable in Raman-active molecules and solids.
Indeed, our scheme for {producing hybrid photon-phonon entanglement} is agnostic to sample details and is passively phase-stabilized, while our two-color pump-probe technique can address Raman-active vibrations irrespective of any  polarization selection rules -- all of which differ from earlier work \cite{lee2011science}. 
Our results demonstrate the strongest form of quantum correlations and is thus a powerful generalization of techniques deployed in atomic physics to study the decoherence of entanglement \cite{de2006direct}.

\begin{figure*}[t!]
	\centering
	\includegraphics[width=1\textwidth]{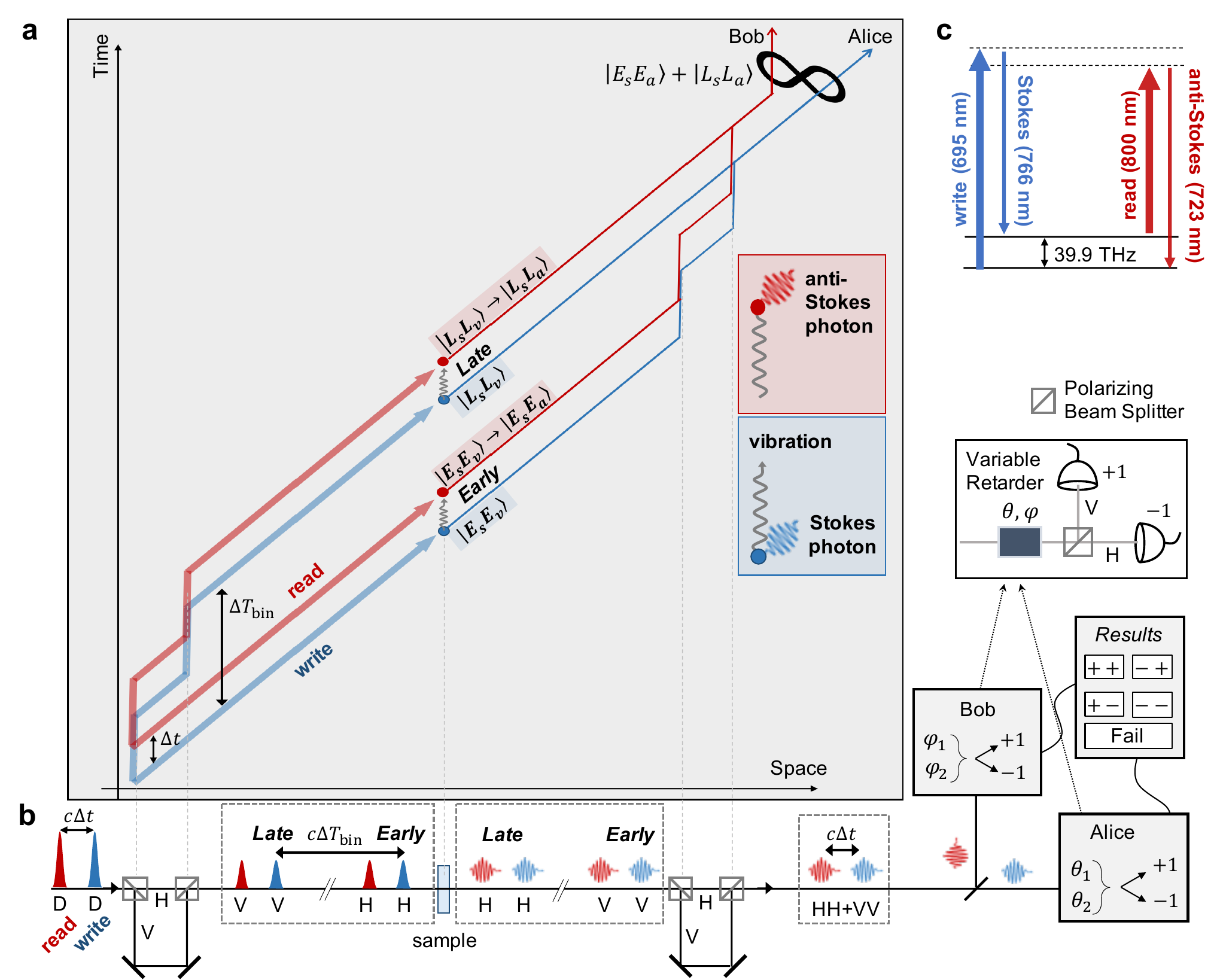}
	\caption{\label{fig:SimplifiedSetup} \textbf{Conceptual scheme and simplified experimental layout.} 
	\textbf{a}, Space-time diagram representation of the time-bin entanglement procedure.
	\textbf{b}, Corresponding experimental implementation unfolded in space along the horizontal axis (see Suppl. Material, Section~1 for details). Contents of the dashed boxes illustrate the time sequence and polarization of the excitation pulses 
	(Gaussian wavepackets) and Raman-scattered photons (wavy arrows),  
	during a single repetition of the experiment. The polarization states are denoted by D (diagonal), H (horizontal) and V (vertical). Note that in our geometry the polarization of Raman scattered photons is orthogonal to that of the incoming pulses. 
	The vertical dashed lines  in panel \textbf{a} correspond to different points in space along the setup. \textbf{c}, Energy diagram of the relevant Raman  interactions, showing the center wavelengths used in the experiment.
}
\end{figure*}


\section*{Materials and Methods}
The inelastic scattering of light off an internal vibrational mode  --- vibrational Raman scattering 
--- is analogous to the radiation-pressure interaction between light and a mechanically compliant mirror \cite{roelli2016}.  
Specifically, the Raman interaction consists of two processes.
In the Stokes process, a quantum of vibrational energy $\hbar \Omega_v$ (a phonon) is created together with a quantum of electromagnetic energy $\hbar \omega_s$ (a Stokes photon); in the anti-Stokes process 
a phonon is annihilated while an anti-Stokes photon is created at angular frequency $\omega_{a}$. 
Energy conservation demands that $\omega_{s,a} \pm \Omega_v=\omega_\text{in}$ respectively,
where $\omega_\text{in}$ is the frequency of the incoming photon. 

In our experiment, a diamond sample --- grown along the [100] direction by high-pressure high-temperature method, about 
300 $\mathrm{\mu m}$ thick and polished on both faces along the (100) crystallographic plane  --- is excited with femtosecond pulses from a mode-locked laser through a pair of high numerical aperture objectives (NA=0.8). 
(The effective length over which the Raman interaction takes place is of the order of 2 $\mathrm{\mu m}$.)
Since the pulses are shorter than the coherence time of the Raman-active vibration, but longer than its oscillation period, there exists perfect time correlation between the generation (resp. annihilation) of a vibrational excitation and the production of a Stokes (resp. anti-Stokes) photon.
In the following, we show how to leverage this time correlation to generate time-bin entanglement \cite{Marci02} between two effective photonic qubits that reveal properties of the mediating phonon mode, and quantify the strength of the quantum correlations using the CHSH form of the Bell inequality \cite{freedman1972}.

The scheme (Fig.~\ref{fig:SimplifiedSetup}) starts when a pair of laser pulses, labeled ``write'' and ``read'' impinging on the sample. Each is a classical wavepacket with $\sim 10^8$ photons per pulse. Their central frequencies are independently tunable, which allows spectral filtering of the Stokes field generated by the write pulse and the anti-Stokes field generated by the read pulse, which are sent to separate detection apparatuses. The delay between them, $\Delta t$, is adjustable to probe the decoherence of the {vibrational mode}. 
Each pulse passes through an unbalanced Mach-Zehnder interferometer and is split in two temporal modes separated by $\Delta T_\text{bin} \gg \Delta t$, which we label the ``early'' and ``late'' time bins. 
$\Delta T_\text{bin} \simeq 3$~ns is chosen to be much longer than the expected vibrational coherence time, which ensures that there can be no quantum-coherent interaction between the two time bins mediated by the vibrational mode.

At room temperature, the thermal state of the vibrational mode \cite{tarrago19} (at 39.9~THz) as a mean occupancy $1.5 \times 10^{-3}$. 
The initial state of the vibration in the two time bins is therefore very well approximated by the ground state $|0_v\rangle \equiv |0_{v,E} \rangle \otimes |0_{v,L} \rangle$, where the subscripts $E$ and $L$ stand for the early and late time bins, respectively.
The Stokes ($s$) and anti-Stokes ($a$) fields are also in the vacuum state at the start of the experiment, denoted by $|0_s\rangle \equiv |0_{s,E} \rangle \otimes |0_{s,L} \rangle$ and $|0_a\rangle \equiv |0_{a,E} \rangle \otimes |0_{a,L} \rangle$.

The interaction of the write pulse (split into the two time bins)  with the vibrational mode generates a two-mode squeezed state of the Stokes and vibrational fields \cite{tarrago19} in each time bin.
A read pulse delayed by $\Delta t$  (also split into the two time bins) maps the vibrational state in the respective
time bins onto its anti-Stokes sideband.

Since we perform the experiment in the regime of very low Stokes scattering probability and post-select the outcomes where exactly one Stokes photon and one anti-Stokes photon were detected (see SM for the treatment of triple coincidence), our scheme can be described in a sub-space of the full Hilbert space that contains one vibrational excitation only, shared by the early and late time bin. 
We therefore introduce the shortened notation $|E_v \rangle \equiv \hat{v}^\dagger_E |0_v\rangle;  |L_v \rangle \equiv \hat{v}^\dagger_L |0_v\rangle$ for the single phonon states (here $\hat{v}^\dagger$ is the phonon creation operator),  
and $|E_s \rangle \equiv \hat{s}^\dagger_E |0_s\rangle$; $ |L_s \rangle \equiv \hat{s}^\dagger_L |0_s\rangle$ for the Stokes single photon states (here $\hat{s}^\dagger$ is the Stokes photon creation operator). 
Conditioned on the detection of a single Stokes photon, the hybrid light--vibrational state {can be} written in the basis $\{ |E_s \rangle , |L_s \rangle \}\otimes \{ |E_v \rangle , |L_v \rangle \} = \{ |E_s\rangle \otimes |E_v\rangle, |E_s\rangle \otimes |L_v\rangle, |L_s\rangle \otimes |E_v\rangle, |L_s\rangle \otimes |L_v\rangle \}$. 
In this sense, we can speak of vibrational and photonic qubits encoded in the time bin basis. 

Within each time bin, the read pulse implements (with a small probability $\sim$0.1\%) the map  $ |E_s, E_v \rangle \rightarrow |E_s, E_a \rangle$ and $ |L_s, L_v \rangle \rightarrow |L_s, L_a \rangle$, where we have defined $|E_a \rangle \equiv \hat{a}^\dagger_E |0_a\rangle$ and $  |L_a \rangle \equiv \hat{a}^\dagger_L |0_a\rangle$ (here $\hat{a}^\dagger_{E,L}$ are the creation operators for the anti-Stokes photon in each time bin). 
Detection of an anti-Stokes photon in coincidence with a Stokes photon from the write pulse heralds that the time bin qubit was successfully mapped onto an anti-Stokes photonic qubit.

By passing the Stokes and anti-Stokes photons through an unbalanced interferometer identical to the one used on the excitation path (Fig.~\ref{fig:SimplifiedSetup}b, and SM
), ``which-time'' information is erased. Moreover, the use of polarizing beam splitters in the interferometer maps
the time-bin-encoded Stokes and anti-Stokes photonic qubits onto polarization-encoded qubits after they are temporally overlapped, $|E_s, E_a \rangle \rightarrow |V_s, V_a \rangle$ and $|L_s, L_a \rangle \rightarrow |H_s, H_a \rangle$ where $H$ and $V$ refer to two orthogonal polarizations of the same temporal mode.
We thus prepare the heralded Bell correlated state
\begin{equation}
 |\psi_{s,a}\rangle = \frac{1}{\sqrt{2}} \left(|V_s, V_a \rangle - e^{i{\phi}}| H_s, H_a\rangle \right)
 \label{eq:Bell_state_a}
\end{equation}
where the phase ${\phi}$ is the sum of the phases acquired by the Stokes and anti-Stokes photons coming from the late time bin, with respect to the early time bin 
(the apparatus is set to realize ${\phi}=0$). 
As detailed in Fig.~S1, the experiment is passively phase-stable by design.

\begin{figure}[t!]
	\centering
	\includegraphics[width=0.66\textwidth]{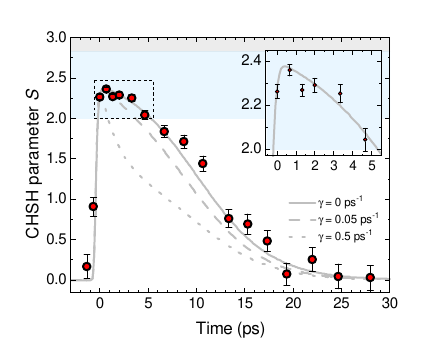}
	\caption{\label{fig:CHSH} \textbf{Time-resolved photon-phonon Bell correlations.}
	The CHSH parameter $S$ (\cref{S_parameter}) as a function of write -- read delay $\Delta t${, with a zoom near $\Delta t = 0$ as an inset}. Full circles are experimental data, while error bars are computed from a Monte Carlo simulation {(see SM, Section~2 for details)}. The solid gray line is obtained from the model with zero pure dephasing and no other free parameters, while dashed lines illustrate the impact of two different non-zero values. The blue region, demarcated by $2 < |S| \leq 2\sqrt{2}$, certifies Bell correlations, while the gray region above it is forbidden for non-superluminal theories.
}
\end{figure}

In order to prove Bell correlations mediated by the room-temperature macroscopic vibration, we send the Stokes and anti-Stokes signals to two independent measurement apparatus labelled Alice and Bob, respectively, who perform local rotations of the Stokes and anti-Stokes states before making a projective measurement in the two-dimensional basis $\{ |V_s \rangle , |H_s \rangle \}$ and $\{ |V_a \rangle , |H_a \rangle \}$, respectively. 
Each party will obtain one of two outcomes, which we label ``$+$'' or ``$-$''.
The number of coincident events where Alice obtains the outcome $x\in \{+,-\}$ and Bob obtains the outcome $y\in \{+,-\}$
is denoted $n_{xy}$.
We then define the
normalized correlation parameter 
\begin{equation}\label{correlation_parameter}
	E_{\theta,\varphi} = \frac{n_{++}+n_{--}-n_{+-}-n_{-+}}{n_{++}+n_{--}+n_{+-}+n_{-+}}
\end{equation}
where the angles $\theta$ and $\varphi$ label the rotations 
that Alice and Bob respectively perform on their qubits before the measurement.
It is defined in such a way that fully correlated events for a given pair of rotation angles $\{\theta,\varphi\}$ yield $E_{\theta,\varphi}=1$ while perfectly anti-correlated events yield $E_{\theta,\varphi}=-1$. 
In fact the CHSH parameter \cite{clauser1969proposed},
\begin{equation}\label{S_parameter}
	S = E_{\theta_1,\varphi_1}+E_{\theta_2,\varphi_2}+E_{\theta_1,\varphi_2}-E_{\theta_2,\varphi_1}
\end{equation}
certifies Bell correlations when $|S|>2$.
In particular, for our scenario, where we target the Bell correlated state eq.~(\ref{eq:Bell_state_a}), a maximal violation is expected for $\{\theta_1,\theta_2\} = \{0, \frac{\pi}{2}\}$ and $\{\varphi_{1},\varphi_{2}\} = \{-\frac{\pi}{4},\frac{\pi}{4}\}$.



\section*{Results}
\subsection*{Observation of Bell correlations}
\Cref{fig:CHSH} shows the CHSH parameter (\cref{S_parameter}) measured for a varying write--read delay.
Our data demonstrates a clear violation of the Bell inequality (whose classical bound is marked as the white region) which persists for more than 5~ps\modif{, about 50 times longer that the write and read pulse duration. While this timescale is consistent with the phonon lifetime in diamond, the dynamics of Bell correlations in fact strongly depends on experimental noise and non-idealities, as explained in SM, Sections 4-6.
} 
At a time delay of $0.66$~ps, for which there is vanishing temporal overlap of the write and read pulses within the sample and correlations are only mediated by the vibration, we measure $S=2.360 \pm 0.025$. 
This confirms Bell correlations mediated by the vibration that acts as a room-temperature quantum memory \cite{england2013,england2015,fisher2016a,bustard2017,fisher2017}.

\begin{figure}[t!]
	\centering
	\includegraphics[width=0.6\textwidth]{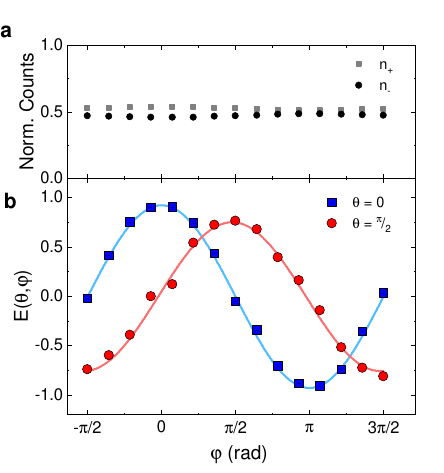}
	\caption{\label{fig:TPI} \textbf{Phonon-mediated two-photon interference.}
	\textbf{a}, Normalized single-photon count rates on the two anti-Stokes detectors as a function of Bob's rotation angle $\varphi$. The ideal marginal state is the statistical mixture $p |E_a\rangle \langle E_a|-(1-p) |L_a\rangle \langle L_a|$, with $p = \tfrac{1}{2}$; data is consistent with $|p - \tfrac{1}{2}| = 0.027$. Error bars are several times smaller than symbol size. 
	\textbf{b}, Two-photon interferences in the Stokes -- anti-Stokes coincidence rate as a function of Bob's rotation angle $\varphi$. The normalized correlation parameter $E(\theta,\varphi)$ (eq.~\ref{correlation_parameter}) is plotted for two fixed angles $\theta=0$ (blue squares) and $\theta= \pi/2$ (red circles) for Alice's rotation on the Stokes state, at a fixed write -- read delay of $\Delta t = 0.66$~ps. Experimental data are represented by full symbols (error bars are smaller than symbol size); solid lines are fitting curves to extract the visibility (see SM for details).
}
\end{figure}

A detailed analysis of the event statistics (see SM, Section~6) enables us to make a more precise claim concerning the violation of the Bell inequality \cite{bancal2018}, without assuming that our data is independent and identically 
distributed. 
From this analysis, we can claim with a confidence level of $1\, - \, 6 \times 10^{-7}$ that the post-selected Stokes -- anti-Stokes state features Bell correlations with a minimum value of the CHSH parameter  $S_\text{min} = 2.23$.

Note that we rely on the fair sampling assumption \cite{Berry10} since the overall detection efficiency in our experiment is not high enough to test a Bell inequality without post-selection of events where at least one detector clicks on each side (Alice and Bob). 
However, it can be shown \cite{orsucci2019} that when all detectors are equally efficient -- a condition well 
approximated in our experiment -- the post-selected data is faithful to that from an ideal experiment where lossless 
devices measure a state obtained by quantum filtering the actual Stokes — anti-Stokes state. 
By reporting a CHSH value higher than 2, we show that this filtered state is Bell-correlated.

To gain further insight into the nature of the Bell correlated state prepared in the experiment, and the reasons why the quantum bound ($\vert S \vert =2\sqrt{2}$) is not saturated, we perform further measurements.
\Cref{fig:TPI}a shows the one-photon counts as Bob's analysis angle is rotated. 
For an ideal Bell state, the marginal is maximally mixed, and should lead to no dependence of the one-photon counts
on the analysis angle. The observed data is consistent with a deviation from a maximal mixture by $2.7\%$.

\Cref{fig:TPI}b shows two-photon interference for various settings of Bob's measurement angle 
for two fixed values of Alice's measurement angle, $\theta= 0 \; , \; \pi/2$, and a fixed write-read delay of 0.66 ps.
The interference is consistent with a model (see SM, Section~3) where the
Stokes interaction creates a two-mode light-vibration squeezed state, and that  anti-Stokes scattering implements a beam-splitter interaction \cite{tarrago19}.
  
The curve for the setting $\theta=0$ (Fig.~\ref{fig:TPI}b blue trace) reveals how accurately we can prepare and distinguish the two states $|E_s,E_a \rangle$ and $|L_s,L_a \rangle$. 
At a given delay, the visibility has an upper limit related to the strength of Stokes -- anti-Stokes photon number correlations, $V_\text{max}=\frac{g^{(2)}_{s,a}-1}{g^{(2)}_{s,a}+1}$ \cite{de2006direct}, where $g^{(2)}_{s,a}$ is the normalized second-order cross-correlation \cite{anderson2018} (see SM). 
The value extracted from the fit is $V_{\theta=0}=93\pm 1 \%$, in agreement with the independently measured value of $g^{(2)}_{s,a}(0)=25$, showing that the signal-to-noise ratio in the cross-correlation is indeed the limiting factor for the visibility in this setting. This visibility could be improved by reducing the power of the write beam (to decrease the probability of creating multiple Stokes-phonon pairs in one pulse) and that of the read beam (to reduce the noise from degenerate four-wave mixing).
Note that due to the small interaction length ($\sim 2~\mu$m), phase matching is not a relevant concern.

The coincidence curve for $\theta=\frac{\pi}{2}$ (Fig.~\ref{fig:TPI}b red trace) corresponds to a rotated measurement basis for Alice and is sensitive to the fluctuations of the phase $\phi$ in the superposition of eq.~(\ref{eq:Bell_state_a}).
To accomodate this possibility, we model the relative phase $\phi$ 
in eq.~(\ref{eq:Bell_state_a}) to be distributed as a zero-mean Gaussian random variable with variance $\sigma$ (see SM). 
We extract a visibility  $V_{\theta=\pi/2}=76\%$ from the fit to the experimental data, which is reproduced by the model for a standard deviation $\sigma=0.31$~rad (equivalent to a $\pm 0.18$~fs timing uncertainty maintained over $\sim 4$ minutes). 

Ultimately, we are able to predict all measured quantities from independently characterised parameters, namely the Raman scattering probability, the overall Raman signal detection efficiency, and the dark count rate of the detectors (see SM, Section 4).

\subsection*{Decoherence dynamics of the phonon mode}
From the temporal behavior of the CHSH parameter we can extract the rate of pure dephasing of the vibrational mode mediating the Bell correlations.
In the absence of pure dephasing, the CHSH parameter decays with the collective vibrational mode. 
Pure dephasing, in contrast, scrambles the phase $\phi$ of the superposition in state~(\ref{eq:Bell_state_a}).
We model it as a random-walk of the phase at the characteristic time scale $\gamma^{-1}$, so that the standard deviation 
of the phase $\phi$ increases with the write--read delay (in addition to technical fluctuations) as 
$\sigma=\sqrt{\gamma\,\Delta t}$  (see SM, Section~3.6).
The model is plotted against the data of Fig.~\ref{fig:CHSH} (solid line), and the best agreement with the data is obtained with a pure dephasing rate identically null (other pure dephasing rates are plotted for comparison), consistent with previous measurements of the coherence time of a single vibrational mode in diamond using transient coherent ultrafast phonon spectroscopy \cite{waldermann2008}.


\section*{Discussion}
For the first time, we have produced Bell correlations between two photons through their interaction with a common Raman-active phonon at room temperature, and probed their decay with sub-picosecond resolution.
Remarkably, our data show that Bell correlations are preserved for more than $200$ oscillation periods at room temperature, evidencing a mechanical coherence time in par with the state-of-the-art for microfabricated resonators under high vaccum \cite{ghad18}.
Optical phonons in diamond indeed exhibit a room temperature ``Q-frequency product'' of $\sim 4 \times 10^{16}$ Hz, making them attractive resonators for ultrafast quantum technologies.

\modif{
Such highly coherent vibrational modes, together with the toolset of time-resolved single photon Raman spectroscopy that we have demonstrated here, should allow to entangle two vibrational qubits via entanglement swapping \cite{zukowski1993}, or to perform optomechanical conversion between photonic qubits at different frequencies \cite{hill2012}, among other possible applications. Much longer vibrational coherence times could be achieved with ensembles of molecules that are decoupled from the phonon bath by surface engineering \cite{chen2018e} or optical trapping and cooling \cite{kondov2019}. Besides, molecules in the gas phase exhibit more complex mechanical degrees of freedom, including rotational and rovibrational modes \cite{koch2019}, with increased coherence time and rich opportunities for quantum information processing \cite{albert2020}. In the future, our scheme could be applied to individual molecules free of heterogeneous broadening using the enhancement of light-vibration coupling offered by electronic resonances \cite{maser2016}, plasmonic nanocavities \cite{yampolsky2014} or optical microcavities \cite{Riedel19}.”

}

In addition to being a benchmark for the robust generation of optomechanical Bell correlations at room temperature, our work suggests a new class of techniques able to probe the role of phonon-mediated entanglement in quantum technologies \cite{aspelmeyer2014}, chemistry \cite{halpin2014}, or even biology \cite{duan2017}. 

\section*{Acknowledgments}
The authors thank T. J. Kippenberg and J-P. Brantut for valuable discussion; and Prof Angelo Geraci, Nicola Lusardi and Fabio Garzetti for providing the custom FPGA-based correlation electronics. 
This works was funded by the Swiss National Science Foundation (SNSF) (project number PP00P2-170684), and 
the European Research Council's (ERC) Horizon 2020 research and innovation programme (grant agreement No. 820196).
N.S. acknowledges funding by the Swiss National Science Foundation (SNSF), through the Grant PP00P2-179109, by the Army Research Laboratory Center for Distributed Quantum Information via the project SciNet and from the European Union’s Horizon 2020 research and innovation programme under grant agreement No 820445 and project name Quantum Internet Alliance. 

\subsection*{Author contributions} S.T.V., N.S. and C.G. designed the experiment; S.T.V. performed the measurements and analysed the data; all authors discussed and interpreted the results and wrote the manuscript.

\subsection*{Competing interests} The authors declare that they have no competing interests.

\subsection*{Data and materials availability} The data related to this paper is available via a Zenodo repository at \url{https://doi.org/10.5281/zenodo.4084706}. Additional data related to this paper may be requested from the authors. 

\section*{Supplementary Material}

Supplementary material containing details of the experimental setup, data analysis and theoretical model, is available online. It contains references \cite{chen2018b,sekatski2012,bancal2018}

\bibliography{BellCorrelations}

\begin{thebibliography}{10}

\bibitem{bell1964einstein}
J.~S. Bell, ``On the einstein podolsky rosen paradox,'' {\em Physics}, vol.~1,
  no.~3, p.~195, 1964.

\bibitem{clauser1969proposed}
J.~F. Clauser, M.~A. Horne, A.~Shimony, and R.~A. Holt, ``Proposed experiment
  to test local hidden-variable theories,'' {\em Phys. Rev. Lett.}, vol.~23,
  no.~15, p.~880, 1969.

\bibitem{Brunner14}
N.~Brunner, D.~Cavalcanti, S.~Pironio, V.~Scarani, and S.~Wehner, ``Bell
  nonlocality,'' {\em Reviews of Modern Physics}, vol.~86, p.~419, 2014.

\bibitem{freedman1972}
S.~J. Freedman and J.~F. Clauser, ``Experimental test of local hidden-variable
  theories,'' {\em Phys. Rev. Lett.}, vol.~28, no.~14, p.~938, 1972.

\bibitem{AspRog81}
A.~Aspect, P.~Grangier, and G.~Roger, ``Experimental tests of realistic local
  theories via bell's theorem,'' {\em Phys. Rev. Lett.}, vol.~47, p.~460, 1981.

\bibitem{chou2007}
C.-W. Chou, J.~Laurat, H.~Deng, K.~S. Choi, H.~de~Riedmatten, D.~Felinto, and
  H.~J. Kimble, ``Functional {{Quantum Nodes}} for {{Entanglement
  Distribution}} over {{Scalable Quantum Networks}},'' {\em Science}, vol.~316,
  pp.~1316--1320, June 2007.

\bibitem{hofmann2012}
J.~Hofmann, M.~Krug, N.~Ortegel, L.~G{\'e}rard, M.~Weber, W.~Rosenfeld, and
  H.~Weinfurter, ``Heralded {{Entanglement Between Widely Separated Atoms}},''
  {\em Science}, vol.~337, pp.~72--75, July 2012.

\bibitem{blatt_entangled_2008}
R.~Blatt and D.~Wineland, ``Entangled states of trapped atomic ions,'' {\em
  Nature}, vol.~453, p.~1008, 2008.

\bibitem{matsukevich2005}
D.~N. Matsukevich, T.~Chaneli{\`e}re, M.~Bhattacharya, S.-Y. Lan, S.~D.
  Jenkins, T.~A.~B. Kennedy, and A.~Kuzmich, ``Entanglement of a {{Photon}} and
  a {{Collective Atomic Excitation}},'' {\em Phys. Rev. Lett.}, vol.~95,
  p.~040405, July 2005.

\bibitem{kuzm13}
L.~Li, O.~Dudin, and A.~Kuzmich, ``Entanglement between light and an optical
  atomic excitation,'' {\em Nature}, vol.~498, p.~466, 2013.

\bibitem{engelsen17}
N.~J. Engelsen, R.~Krishnakumar, O.~Hosten, and M.~Kasevich, ``Bell
  correlations in spin-squeezed states of 500 000 atoms,'' {\em Phys. Rev.
  Lett.}, vol.~118, p.~140401, 2017.

\bibitem{ansmann2009violation}
M.~Ansmann, H.~Wang, R.~C. Bialczak, M.~Hofheinz, E.~Lucero, M.~Neeley,
  A.~O'Connell, D.~Sank, M.~Weides, J.~Wenner, {\em et~al.}, ``Violation of
  bell's inequality in josephson phase qubits,'' {\em Nature}, vol.~461,
  no.~7263, p.~504, 2009.

\bibitem{carlo10}
L.~DiCarlo, M.~D. Reed, L.~Sun, B.~R. Johnson, J.~M. Chow, J.~M. Gambetta,
  L.~Frunzio, S.~M. Girvin, M.~H. Devoret, and R.~J. Schoelkopf, ``Preparation
  and measurement of three-qubit entanglement in a superconducting circuit,''
  {\em Nature}, vol.~467, p.~574, 2010.

\bibitem{morello16}
J.~P. Dehollain, S.~Simmons, J.~Muhonen, R.~Kalra, A.~Laucht, F.~Hudson,
  K.~Itoh, D.~Jameison, J.~McCallum, A.~Dzurak, and A.~Morello, ``Bell's
  inequality violation with spins in silicon,'' {\em Nat. Nano.}, vol.~11,
  p.~242, 2016.

\bibitem{hensen2015loophole}
B.~Hensen, H.~Bernien, A.~E. Dr{\'e}au, A.~Reiserer, N.~Kalb, M.~S. Blok,
  J.~Ruitenberg, R.~F. Vermeulen, R.~N. Schouten, C.~Abell{\'a}n, {\em et~al.},
  ``Loophole-free bell inequality violation using electron spins separated by
  1.3 kilometres,'' {\em Nature}, vol.~526, no.~7575, p.~682, 2015.

\bibitem{marinkovic2018}
I.~Marinkovi{\'c}, A.~Wallucks, R.~Riedinger, S.~Hong, M.~Aspelmeyer, and
  S.~Gr{\"o}blacher, ``Optomechanical bell test,'' {\em Phys. Rev. Lett.},
  vol.~121, no.~22, p.~220404, 2018.

\bibitem{lee2011science}
K.~C. Lee, M.~R. Sprague, B.~J. Sussman, J.~Nunn, N.~K. Langford, X.-M. Jin,
  T.~Champion, P.~Michelberger, K.~F. Reim, D.~England, {\em et~al.},
  ``Entangling macroscopic diamonds at room temperature,'' {\em Science},
  vol.~334, no.~6060, pp.~1253--1256, 2011.

\bibitem{jorio2015}
A.~Jorio, M.~Kasperczyk, N.~Clark, E.~Neu, P.~Maletinsky, A.~Vijayaraghavan,
  and L.~Novotny, ``Stokes and anti-{{Stokes Raman}} spectra of the high-energy
  {{C}}\textendash{{C}} stretching modes in graphene and diamond,'' {\em Phys.
  Status Solidi B}, vol.~252, pp.~2380--2384, Nov. 2015.

\bibitem{kasperczyk2015}
M.~Kasperczyk, A.~Jorio, E.~Neu, P.~Maletinsky, and L.~Novotny,
  ``Stokes\textendash{}anti-{{Stokes}} correlations in diamond,'' {\em Optics
  Letters}, vol.~40, p.~2393, May 2015.

\bibitem{england2016}
D.~G. England, K.~A.~G. Fisher, J.-P.~W. MacLean, P.~J. Bustard, K.~Heshami,
  K.~J. Resch, and B.~J. Sussman, ``Phonon-{{Mediated Nonclassical
  Interference}} in {{Diamond}},'' {\em Phys. Rev. Lett.}, vol.~117, p.~073603,
  Aug. 2016.

\bibitem{hou2016}
P.-Y. Hou, Y.-Y. Huang, X.-X. Yuan, X.-Y. Chang, C.~Zu, L.~He, and L.-M. Duan,
  ``Quantum teleportation from light beams to vibrational states of a
  macroscopic diamond,'' {\em Nature Communications}, vol.~7, p.~11736, May
  2016.

\bibitem{anderson2018}
M.~D. Anderson, S.~T. Velez, K.~Seibold, H.~Flayac, V.~Savona, N.~Sangouard,
  and C.~Galland, ``Two-color pump-probe measurement of photonic quantum
  correlations mediated by a single phonon,'' {\em Phys. Rev. Lett.}, vol.~120,
  no.~23, p.~233601, 2018.

\bibitem{bustard2015}
P.~J. Bustard, J.~Erskine, D.~G. England, J.~Nunn, P.~Hockett, R.~Lausten,
  M.~Spanner, and B.~J. Sussman, ``Nonclassical correlations between
  terahertz-bandwidth photons mediated by rotational quanta in hydrogen
  molecules,'' {\em Opt. Lett., OL}, vol.~40, pp.~922--925, Mar. 2015.

\bibitem{kasperczyk2016}
M.~Kasperczyk, F.~S. {de Aguiar J{\'u}nior}, C.~Rabelo, A.~Saraiva, M.~F.
  Santos, L.~Novotny, and A.~Jorio, ``Temporal {{Quantum Correlations}} in
  {{Inelastic Light Scattering}} from {{Water}},'' {\em Phys. Rev. Lett.},
  vol.~117, p.~243603, Dec. 2016.

\bibitem{saraiva2017}
A.~Saraiva, F.~S. d.~A. J{\'u}nior, R.~{de Melo e Souza}, A.~P. Pena, C.~H.
  Monken, M.~F. Santos, B.~Koiller, and A.~Jorio, ``Photonic {{Counterparts}}
  of {{Cooper Pairs}},'' {\em Phys. Rev. Lett.}, vol.~119, p.~193603, Nov.
  2017.

\bibitem{tarrago19}
S.~T. Velez, K.~Seibold, N.~Kipfer, M.~D. Anderson, V.~Sudhir, and C.~Galland,
  ``Preparation and decay of a single quantum of vibration at ambient
  conditions,'' {\em Phys. Rev. X}, vol.~9, p.~041007, 2019.

\bibitem{de2006direct}
H.~De~Riedmatten, J.~Laurat, C.-W. Chou, E.~Schomburg, D.~Felinto, and H.~J.
  Kimble, ``Direct measurement of decoherence for entanglement between a photon
  and stored atomic excitation,'' {\em Phys. Rev. Lett.}, vol.~97, no.~11,
  p.~113603, 2006.

\bibitem{roelli2016}
P.~Roelli, C.~Galland, N.~Piro, and T.~J. Kippenberg, ``Molecular cavity
  optomechanics as a theory of plasmon-enhanced raman scattering,'' {\em Nat.
  Nano.}, vol.~11, no.~2, p.~164, 2016.

\bibitem{Marci02}
I.~Marcikic, H.~de~Riedmatten, W.~Tittel, V.~Scarani, H.~Zbinden, and N.~Gisin,
  ``Time-bin entangled qubits for quantum communication created by femtosecond
  pulses,'' {\em Phys. Rev. A}, vol.~66, p.~062308, Dec. 2002.

\bibitem{england2013}
D.~G. England, P.~J. Bustard, J.~Nunn, R.~Lausten, and B.~J. Sussman, ``From
  {{Photons}} to {{Phonons}} and {{Back}}: {{A THz Optical Memory}} in
  {{Diamond}},'' {\em Phys. Rev. Lett.}, vol.~111, p.~243601, Dec. 2013.

\bibitem{england2015}
D.~G. England, K.~Fisher, J.-P.~W. MacLean, P.~J. Bustard, R.~Lausten, K.~J.
  Resch, and B.~J. Sussman, ``Storage and {{Retrieval}} of {{THz}}-{{Bandwidth
  Single Photons Using}} a {{Room}}-{{Temperature Diamond Quantum Memory}},''
  {\em Phys. Rev. Lett.}, vol.~114, p.~053602, Feb. 2015.

\bibitem{fisher2016a}
K.~A.~G. Fisher, D.~G. England, J.-P.~W. MacLean, P.~J. Bustard, K.~J. Resch,
  and B.~J. Sussman, ``Frequency and bandwidth conversion of single photons in
  a room-temperature diamond quantum memory,'' {\em Nat. Commun.}, vol.~7,
  p.~11200, Apr. 2016.

\bibitem{bustard2017}
P.~J. Bustard, D.~G. England, K.~Heshami, C.~Kupchak, and B.~J. Sussman,
  ``Quantum frequency conversion with ultra-broadband tuning in a {{Raman}}
  memory,'' {\em Phys. Rev. A}, vol.~95, p.~053816, May 2017.

\bibitem{fisher2017}
K.~A.~G. Fisher, D.~G. England, J.-P.~W. MacLean, P.~J. Bustard, K.~Heshami,
  K.~J. Resch, and B.~J. Sussman, ``Storage of polarization-entangled
  {{THz}}-bandwidth photons in a diamond quantum memory,'' {\em Phys. Rev. A},
  vol.~96, p.~012324, July 2017.

\bibitem{bancal2018}
J.-D. Bancal, K.~Redeker, P.~Sekatski, W.~Rosenfeld, and N.~Sangouard,
  ``Device-independent certification of an elementary quantum network link,''
  2018.

\bibitem{Berry10}
D.~Berry, H.~Jeong, M.~Stobinska, and T.~C. Ralph, ``Fair-sampling assumption
  is not necessary for testing local realism,'' {\em Phys. Rev. A}, vol.~81,
  p.~012109, 2010.

\bibitem{orsucci2019}
D.~Orsucci, J.-D. Bancal, N.~Sangouard, and P.~Sekatski, ``How post-selection
  affects device-independent claims under the fair sampling assumption,'' {\em
  {Quantum}}, vol.~4, p.~238, Mar. 2020.

\bibitem{waldermann2008}
F.~Waldermann, B.~J. Sussman, J.~Nunn, V.~Lorenz, K.~Lee, K.~Surmacz, K.~Lee,
  D.~Jaksch, I.~Walmsley, P.~Spizziri, {\em et~al.}, ``Measuring phonon
  dephasing with ultrafast pulses using raman spectral interference,'' {\em
  Phys. Rev. B}, vol.~78, no.~15, p.~155201, 2008.

\bibitem{ghad18}
A.~H. Ghadimi, S.~A. Fedorov, N.~J. Engelsen, M.~J. Bereyhi, R.~Schilling,
  D.~J. Wilson, and T.~J. Kippenberg, ``Elastic strain engineering for ultralow
  mechanical dissipation,'' {\em Science}, vol.~360, p.~764, 2018.

\bibitem{zukowski1993}
M.~{\.Z}ukowski, A.~Zeilinger, M.~A. Horne, and A.~K. Ekert,
  ````{{Event}}-ready-detectors'' {{Bell}} experiment via entanglement
  swapping,'' {\em Phys. Rev. Lett.}, vol.~71, pp.~4287--4290, Dec. 1993.

\bibitem{hill2012}
J.~T. Hill, A.~H. {Safavi-Naeini}, J.~Chan, and O.~Painter, ``Coherent optical
  wavelength conversion via cavity optomechanics,'' {\em Nat. Commun.}, vol.~3,
  p.~1196, 2012.

\bibitem{chen2018e}
L.~Chen, J.~A. Lau, D.~Schwarzer, J.~Meyer, V.~B. Verma, and A.~M. Wodtke,
  ``The {{Sommerfeld}} ground-wave limit for a molecule adsorbed at a
  surface,'' {\em Science}, vol.~363, p.~158, 2018.

\bibitem{kondov2019}
S.~S. Kondov, C.-H. Lee, K.~H. Leung, C.~Liedl, I.~Majewska, R.~Moszynski, and
  T.~Zelevinsky, ``Molecular lattice clock with long vibrational coherence,''
  {\em Nat. Phys.}, vol.~15, p.~1118, 2019.

\bibitem{koch2019}
C.~P. Koch, M.~Lemeshko, and D.~Sugny, ``Quantum control of molecular
  rotation,'' {\em Reviews of Modern Physics}, vol.~91, no.~3, p.~035005, 2019.

\bibitem{albert2020}
V.~V. Albert, J.~P. Covey, and J.~Preskill, ``Robust encoding of a qubit in a
  molecule,'' {\em Physical Review X}, vol.~10, no.~3, p.~031050, 2020.

\bibitem{maser2016}
A.~Maser, B.~Gmeiner, T.~Utikal, S.~G{\"o}tzinger, and V.~Sandoghdar,
  ``Few-photon coherent nonlinear optics with a single molecule,'' {\em Nat
  Photon}, vol.~10, pp.~450--453, July 2016.

\bibitem{yampolsky2014}
S.~Yampolsky, D.~A. Fishman, S.~Dey, E.~Hulkko, M.~Banik, E.~O. Potma, and
  V.~A. Apkarian, ``Seeing a single molecule vibrate through time-resolved
  coherent anti-{{Stokes Raman}} scattering,'' {\em Nat Photon}, vol.~8,
  pp.~650--656, Aug. 2014.

\bibitem{Riedel19}
D.~Riedel, S.~Fl\aa{}gan, P.~Maletinsky, and R.~J. Warburton, ``Cavity-enhanced
  raman scattering for in situ alignment and characterization of solid-state
  microcavities,'' {\em Phys. Rev. Applied}, vol.~13, p.~014036, Jan 2020.

\bibitem{aspelmeyer2014}
M.~Aspelmeyer, T.~J. Kippenberg, and F.~Marquardt, ``Cavity optomechanics,''
  {\em Rev. Mod. Phys.}, vol.~86, pp.~1391--1452, Dec. 2014.

\bibitem{halpin2014}
A.~Halpin, P.~J.~M. Johnson, R.~Tempelaar, R.~S. Murphy, J.~Knoester, T.~L.~C.
  Jansen, and R.~J.~D. Miller, ``Two-dimensional spectroscopy of a molecular
  dimer unveils the effects of vibronic coupling on exciton coherences,'' {\em
  Nat. Chem.}, vol.~6, pp.~196--201, Mar. 2014.

\bibitem{duan2017}
H.-G. Duan, V.~I. Prokhorenko, R.~J. Cogdell, K.~Ashraf, A.~L. Stevens,
  M.~Thorwart, and R.~J.~D. Miller, ``Nature does not rely on long-lived
  electronic quantum coherence for photosynthetic energy transfer,'' {\em Proc
  Natl Acad Sci U S A}, vol.~114, pp.~8493--8498, Aug. 2017.

\bibitem{chen2018b}
X.~Chen, X.~Lu, S.~Dubey, Q.~Yao, S.~Liu, X.~Wang, Q.~Xiong, L.~Zhang, and
  A.~Srivastava, ``Entanglement of single-photons and chiral phonons in
  atomically thin {{WSe}} 2,'' {\em Nat. Phys.}, vol.~15, pp.~221--227, 2019.

\bibitem{sekatski2012}
P.~Sekatski, N.~Sangouard, F.~Bussieres, C.~Clausen, N.~Gisin, and H.~Zbinden,
  ``Detector imperfections in photon-pair source characterization,'' {\em J.
  Phys. B At. Mol. Opt. Phys.}, vol.~45, p.~124016, June 2012.

\end{thebibliography}

\bibliographystyle{ieeetr}

\newpage

\section{Experimental Methods}

\subsection{Description of setup and experimental parameters}

\begin{figure}[h!]
	\centering
	\includegraphics[width=\textwidth]{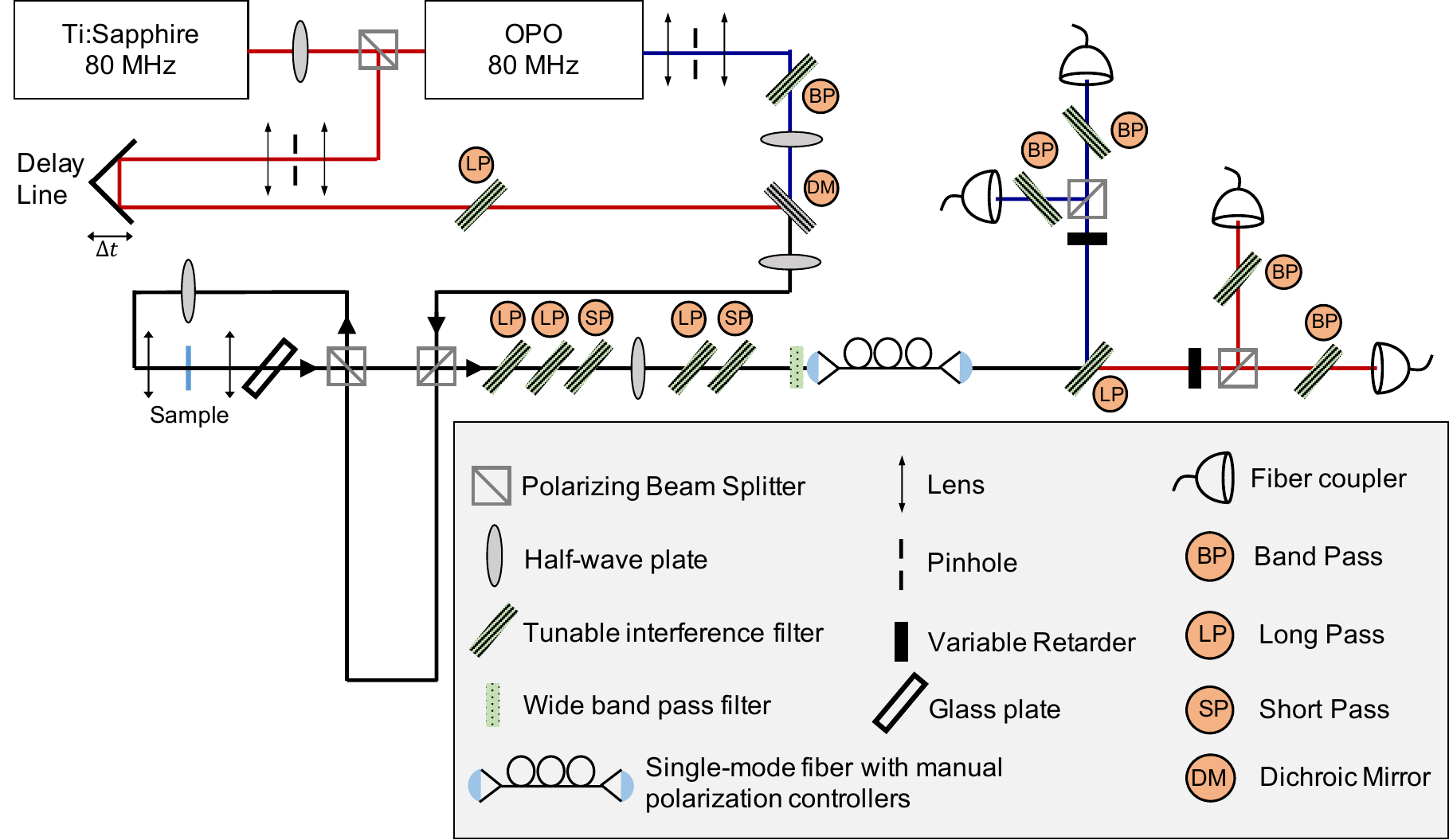}
	\caption{\label{fig:SI_FullSetup} \textbf{Experimental Setup}
}
\end{figure}

The full schematic of the experimental setup is shown in Fig.~\ref{fig:SI_FullSetup}.
We summarize the key experimental parameters in Table \ref{tab:exp_param}.

\begin{table}[h!]
\begin{center}
\begin{tabular}{ |c|c| } 
 \hline
 \textbf{Parameter} & \textbf{Value}  \\ 
 \hline
	Repetition Rate & $80.7$~MHz  \\ 
 \hline
 	Write pulse wavelength & $695$~nm  \\ 
 \hline
   	Stokes wavelength & $766$~nm  \\ 
 \hline
 	Read pulse wavelength & $800$~nm  \\ 
 \hline
   	Anti-Stokes wavelength & $723$~nm  \\ 
 \hline
 	Write pulse energy &  $25$~pJ \\ 
 \hline
 	Read pulse energy &  $248$~pJ \\ 
 \hline 	 
 	Total acquisition time per setting & $4$~min  \\ 
 \hline
 	Average Stokes countrate$^*$ & $35700$~s$^{-1}$   \\ 
 \hline
 	Average anti-Stokes countrate$^*$ &  $1750$~s$^{-1}$\\ 
 \hline
  	Stokes - anti-Stokes coincidence rate$^{* \bullet}$ &  $17$~s$^{-1}$ \\
 \hline  
\end{tabular}
\caption{Summary of relevant experimental parameters. \newline 
$^*$ Calculated using the total countrate of $+$ and $-$ detectors in each detection arm.\newline
$^\bullet$ For a delay $\Delta t = 0.66$~ps.
}\label{tab:exp_param}
\end{center}
\end{table}

\paragraph{Excitation pulses}
A mode-locked Ti:Sapph laser and a synchronously pumped optical parametric oscillator are used to generate the read and write pulses, respectively. The pulse durations are about 100~fs and 200~fs for the Ti:Sapph and OPO, respectively.
The experiment is repeated every 12.5~ns, set by the 80~MHz repetition rate of the laser system.
The linear polarisation of the write and read pulses are first rotated by 45 degrees so that half of their intensity is directed toward the two arms of the unbalanced interferometer, which is constructed with polarising beam splitters.
Light that is vertically polarized travels through the short path, while horizontally polarized light travels through the long path. 
A half-wave plate rotates the polarisation of all pulses by 90 degrees after the interferometer, yielding the pulse distribution shown in Fig.~1b of the main text, where the layout was unfolded and modified for clarity.
The delay between the two arms of the unbalanced interferometer is about $3$~ns (approximately 1~m in free space), orders-of-magnitude longer than the phonon lifetime in the sample ($\approx 4$~ps). 

\paragraph{Which-time information erasing}
The time-bin entangled state is prepared by erasing temporal information about the Raman scattering processes. 
To do so, after the sample the Raman scattered photons are collected in transmission and passed through the same polarisation-selective unbalanced interferometer as the one used to create the two time bins in excitation, but they enter from another input port. 

By suitably rotating the polarisation of the incoming pulses and of the Raman scattered photons, which are related by the symmetry of the vibrational mode under study, we can optimise the likelihood for the Raman photons to temporally overlap after the second interferometer.
This is achieved when Raman scattered photons from the early time-bin are routed in the long arm, and vice-versa. 
Due to the linear polarisation of the Raman scattered fields in our geometry, this likelihood is close to unity (note that for diamond excited along the [100] crystal axis the Raman scattered photons are orthogonally polarised with respect to the pump).
 
In the worst-case scenario where Raman photons are unpolarised, half of them would take the wrong path and remain distinguishable in time. 
Accordingly, the likehood to erase which-time information would drop to one fourth (25\%); but the fidelity of the post-selected entangled state would not be affected. 
Indeed, as long as the time-bin separation is larger than the detector jitter, temporal filtering can be performed to exclude the distinguishable events from analysis. 
It is worth mentioning here that crystals and molecules with different symmetries may allow for the storage of polarisation-encoded vibrational qubits, therefore opening new experimental possibilities to probe photon-phonon entanglement [52]. 

Additionally, using the same physical interferometer twice - first to define the time bins in excitation and then to erase the temporal information carried by the Raman scattered photons - renders our entire setup passively phase-stable, as any fluctuation of the optical path between two arms occurring on a time scale longer than the travel time for light through the setup (which is a few tens of nanoseconds) is cancelled by construction (see detailed layout in SI). 
In this way, we are insensitive to all types of noise causing path fluctuations in a bandwidth of at least $10$~MHz, which encompasses almost all mechanical and thermal instability.

\paragraph{Impact of birefringence}

Special care must be taken to avoid birefringence in the setup, as it would result in a temporal shift between the horizontal (H) and vertical (V) polarisations. Imperfect temporal overlap translates into a mixed state component as opposed to a pure entangled state (see Sec.~\ref{sec:model}). The short duration of the laser pulses means that the overlap must be preserved to well below $100$~fs, and this must be the case for a relatively broad wavelength range of several tens of nm.

Dichcroic mirrors and tunable interference filters in particular have a strong birefringence when the incident angle is non zero. 
We mitigate the birefringence induced delay caused by the dichroic mirror that serves to overlap the write and read pulses on the same spatial mode by preparing both beams in the vertical polarisation, and then using an achromatic half wave plate to rotate the polarisation by 45 degrees before the imbalanced polarised interferometer. 

Also, we must mitigate the deleterious effect of birefringence in the interference filters used to reject the write and read laser beams. 
For this, we use two identical sets (consisting each of a long pass filter to block the write pulse and a short filter to block the read pulse) and place an achromatic half wave plate between them. 
In this way, the Raman scattered light goes through the second sets of filters after its polarisation was rotated by $90$ degrees, so that we ensure that both polarisations are equally delayed even in the presence of birefringence, and thus the temporal and spatial overlap is perfectly maintained.

\paragraph{Time-bin to polarisation qubit mapping}
Since we use polarising beam splitters to route the photons in the short and long path of the unbalanced interferometer, the polarisation is the only degree of freedom that distinguishes between the early and late time bin after the Stokes and anti-Stokes photons are temporally overlapped. 
More specifically, Raman photons originating from the early time are vertically polarised, and those from the late time bin are horizontally polarised. 
 
\paragraph{Detection}

After the laser rejection filters, the Raman signal is spatially filtered by coupling it into a S630-HP single mode fiber (Thorlabs, FC/PC).  
Polarisation control paddles are used to maintain the same linear polarisation before and after the fiber.  
The signal is collimated after the fiber and sent onto  a tunable long pass filter, where the Stokes field is transmitted and the anti-Stokes field is reflected, after which the two fields enter the two detection apparatuses labelled `Alice' and `Bob', respectively. 
The birefringence introduced by this filter  - especially for the reflected beam, which has a very strong wavelength and angle dependence - cannot be easily compensated, and we attribute the main loss of visibility to this element.

At each locations we first use a variable retarder (VR), whose fast axis is rotated by 45 degrees with respect to the vertical, in order to perform the state rotation (see Sec.~\ref{sec:model} for the mathematical formalism). 
Subsequently, a polarising beam splitter (PBS) directs the horizontal (H) and vertical (V) components of the incoming light onto two distinct single photon detectors, implementing thereby a projective measurement in the H/V basis, equivalent to the early/late basis for the time bin qubits. 
After these last PBS, birefringence no longer affects the experiment, and we send the output of each PBS through a tunable bandpass filter centered on the Stokes or anti-Stokes wavelengths, respectively. Finally, we couple each of the four output beams into a multi-mode fiber connected to an avalanche photo diode (APD) operated in Geiger mode, featuring about 50\% detection efficiency and 500~ps timing jitter.

\paragraph{Optimisation}
Before running the experiment we check the two-photon correlations in the $\{\theta=0,\varphi=0\}$ and $\{\theta=\frac{\pi}{2},\varphi=\frac{\pi}{2}\}$ configurations, where $\theta$ (resp. $\varphi$) is the state rotation angle (given by the retardation of the variable retarder) chosen by Alice (resp. Bob). Under ideal conditions we would expect $E(0,0) = E(\frac{\pi}{2},\frac{\pi}{2})$, but we always measure $E(0,0) > E(\frac{\pi}{2},\frac{\pi}{2})$ due to either imperfect alignment or birefringence that was not properly compensated for (see mathematical explanation in Sec.~\ref{sec:model}). As a final step we slightly change the angle of the first long pass filter after the interferometer to maximize the value of $E(\frac{\pi}{2},\frac{\pi}{2})$.


\subsection{Calibration of the Variable Retarders}

The liquid crystal variable retarders (VR) (from ARCoptix) allow us to apply a voltage-dependent delay along one polarisation axis. This axis is set to $45\deg$, allowing us to rotate the polarisation state of each photon in a plane containing the vertical and horizontal states (see Sec.~\ref{sec:model}). 

\begin{figure}[h!]
	\centering
	\includegraphics[width=\textwidth]{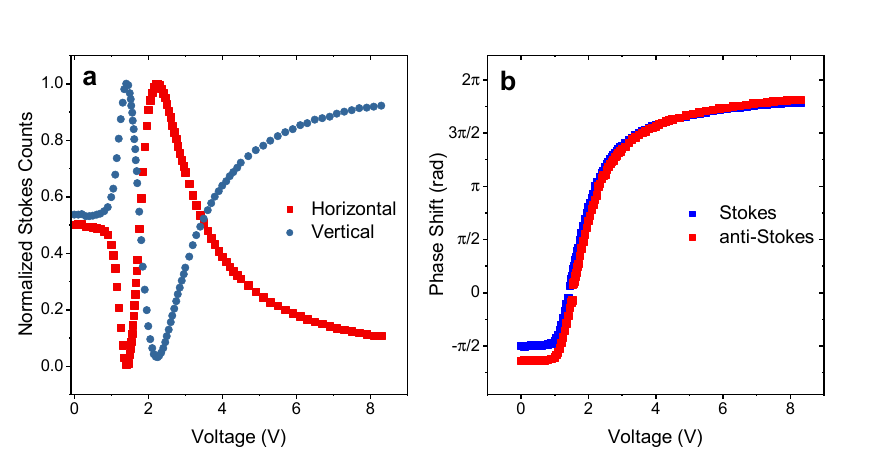}
	\caption{\label{fig:SI_VR_Cal} \textbf{Calibration of variable retarders}
	\textbf{a}, Normalized count rates on the detectors measuring the vertical and horizontal components of the light after the VR, when vertically polarised light is sent into the VR. Data for the Stokes channel only are shown. 
	\textbf{b}, Extracted voltage-dependent phase shifts for the Stokes and anti-Stokes wavelengths.
}
\end{figure}

To avoid any artefact due to the wavelength dependence of the retardation, the calibration of the VRs at Alice and Bob's locations is done with the Stokes or anti-Stokes signals, respectively, by sending vertically polarised light through the VR and  measuring the amount of vertically and horizontally polarised light afterwards using a PBS and two detectors. The phase shift is then $\delta = \arccos(2T-1)$, where $T$ is the normalized count rate in the vertical polarisation detector. Results of this procedure are shown in Fig.~\ref{fig:SI_VR_Cal}.

\section{Data Acquisition Methods}

\subsection{Data Acquisition}
We record the detection events using a custom time-tagging card developed by the Digital Electronics Laboratory at Politecnico di Milano (Prof. Angelo Geraci) with four input channels plus a sync channel. 
Using the sync signal from the mode-locked laser oscillator as a time reference, we define a detection time window that only records photons originating from the early time bin and taking the long path in the second interferometer, or vice versa (from late time bin taking the short path).  We discard all other events. 
Over the acquisition time,  we record the number of single detection events from each detector occurring in the detection window, as well as the number of times multiple detectors clicked during the same window. The number of coincidences between the detectors of Alice and Bob correspond to the $n_{\pm,\pm}$ terms used to calculate $E$.
We obtain the two-photon interference curves of Fig.~2 (main text) at a fixed write-read delay by setting $\theta = 0$ or $(\theta = \frac{\pi}{2})$ and sweeping $\varphi$. 

We ran the Bell tests with the phase setting $\{\theta,\varphi\}$ in the following order: $\{0,\frac{\pi}{4}\}$, $\{0,-\frac{\pi}{4}\}$, $\{\frac{\pi}{2},\frac{\pi}{4}\}$, $\{\frac{\pi}{2},-\frac{\pi}{4}\}$. We measure for 1 minute at each phase setting before changing the delay between write and read pulses, which is moved from negative to positive delay. In order to mitigate systematic errors (drifts in alignment, for example), we repeat the whole measurement sequence four times, for a total of four minutes per measurement setting at each delay setting. For the analysis all the counts of the four measurements with the same setting are added together, and used to compute $E$ and $S$ as explained in the main text.

The value of $g^{(2)}_{s,a}(\Delta t)$ is calculated as
\begin{equation}\label{g2_parameter}
	g^{(2)}_{s,a} = \frac{P(s \cap a)}{P(s)P(a)} = \frac{n_{s \cap a} \cdot R}{n_s \cdot n_a}
\end{equation}
where $R$ is the number of times the experiment was run (the repetition rate of the laser system times the acquisition time), $ n_s $ is the total number of Stokes photons detected (in the appropriate time window), $ n_a $ is the total number of anti-Stokes photons detected, and $n_{s \cap a}$ is the total number of coincidences between Stokes and anti-Stokes photons, i.e. $n_{(s \cap a)} = n_{++}+n_{+-}+n_{-+}+n_{--}$.

\subsection{Error Bars}
The error bars displayed on all experimental plots for the normalised correlation parameter $E$ and CHSH parameter $S$  are calculated using a Monte Carlo approach.
For each measurement we model the probability of each coincidence count $n_{xx}$ as a Poissonian distribution centered on $n_{xx}$. We then pick a random number from each distribution using the Python library NumPy, and use it to calculate $S$ (resp. $E$) for the Bell measurement (resp. visibility measurement). We repeat this process many times in order to obtain a collection of values for $S$ (resp. $E$), and we take the standard deviation of this distribution to be a faithful estimate for the statistical uncertainty of the measurement.

To ensure convergence of the procedure, after each iteration described above, we compare the average and standard deviation of the accumulated values to the results from the previous step. We keep repeating this process until the relative difference between two successive steps is below $10^{-5}$ for both the average and standard deviation.


\section{Theoretical Methods}\label{sec:model}

In this section we explain how we model the experiment in order to obtain the fitting function for the CHSH parameter plotted in Fig.~2 of the main text. We start from the assumptions that i) the photon-phonon state is described by a two-mode squeezed state with phase noise and ii) measurements are done with noisy, non-unit efficiency, non-photon number resolving detectors. We use independent measurements of the squeezing parameter, dark count rates and efficiencies to show that both the result of the cross-correlation measurement and the non-local interference pattern for $\theta = 0$ are consistent with these assumptions. We then use the phase sensitive non-local interference pattern obtained for $\theta = \pi/2$ to evaluate the amount of phase noise. The assumption on the state and the measurements together with the knowledge of the phase noise allows us to predict the time dependence of the CHSH parameter and the effect of pure dephasing on its decay.

\subsection{Modelling of the source and detection devices}

\subsubsection{Source}

We consider a source generating Stokes--anti-Stokes photon pairs according to 

\begin{equation}
	|\psi_{t}\rangle = (1-\mathrm{Th}_g^2)^\frac{1}{2} (1-\mathrm{Th}_{\bar{g}}^2)^\frac{1}{2}
	 e^{\mathrm{Th}_g s^\dagger a^\dagger -
	\mathrm{Th}_{\bar{g}} s^\dagger_\perp a^\dagger_\perp} |\underline{0}\rangle
\end{equation}
where $|\underline{0}\rangle$ denotes the vacuum for all modes; $s$ and $s_\perp$ are bosonic operators corresponding to the two orthogonal modes received by Alice and similarly for Bob. 
In our experiment, these orthogonal modes correspond to the two time bins, which are subsequently converted into orthogonal photon polarisations.
We have used the short notation $\mathrm{Th}_g = \tanh (g)$ where $g$ is the squeezing parameter, is related to the mean photon number in mode $s$ by $\langle \psi_{t} | s^\dagger s | \psi_{t} \rangle = (\sinh g )^2 = (\mathrm{Sh}_g)^2$. In the rest of the text, we also use $\mathrm{Ch}_g = \cosh g$. We specifically consider the symmetric case where $g = \bar{g}$.

\subsubsection{Detector}

We consider photon detectors which do not resolve the photon number. They have an efficiency $\eta$
(overall detection efficiency including all losses from the source to the detector) and a dark count probability $p_{dc}$.
A ``click" event (electric pulse generated by the detector) is then modelled by the POVM [53] 
\begin{equation}
	\hat{D}_s(\eta_s)=1-(1-p_{dc})(1-\eta_s)^{s^\dagger s}
\end{equation}

The subscript (here $s$) specifies the mode which is detected. The dark count probability $p_{dc}$ is the same for all modes and detectors. 
To illustrate the validity of the model, consider the Fock state $|n\rangle$. The probability to get a click is $\langle n | \hat{D}_s(\eta_s) | n\rangle = 1-(1-p_{dc})(1-\eta_s)^n$ ~which equals one minus the probability to lose all the $n$ photons and to get no dark count.

\subsubsection{Choice of measurement settings}

Rotations are possibly performed during detection so that the photons can be measured in
several basis. The detected modes are called $A$ and $A_\perp$ for Alice ($B$ and $B_\perp$ for Bob) and are related to the emission modes by
\begin{equation}
	s = C_\alpha  A + S_\alpha  e^{i \phi_s} A_\perp
\end{equation}
\begin{equation}
	s_\perp = S_\alpha e^{-i \phi_s} A + C_\alpha  A_\perp
\end{equation}
with $C_\alpha  = \cos(\alpha)$ and $S_\alpha  = \sin(\alpha)$ and similarly for Bob.

Note that the angles $\alpha$, $\beta$ are related to the optical phases introduced by the variable retarders in the experiment by $\theta = 2 \alpha$ and $\varphi = 2 \beta$. 
To see why, consider for example the rotation of polarisation by $90^{\circ}$ from vertical to horizontal. 
In this formalism, this corresponds to a rotation angle $\alpha = \pi/2$. 
Experimentally, however, this requires introducing a $\theta=\pi$ phase shift in the variable retarder (whose axis, we recall, is oriented at $45^{\circ}$ w.r.t horizontal and vertical).


\subsubsection{Phase noise}

Consider a mechanism adding a phase which is different for each SPDC process, that is, at a
given run the state can be written as
\begin{equation}\label{eq:Psi_t_phi}
	|\psi_t^\phi \rangle = (1 - \mathrm{Th}_g^2) e^{\mathrm{Th}_g (e^{-i \phi/2} s^\dagger a^\dagger 
	- e^{i \phi/2} s^\dagger_\perp a^\dagger_\perp} |\underline{0}\rangle
\end{equation}
where $\phi$ changes from run to run according to a Gaussian probability distribution $p(\phi) = \frac{1}{\sigma \sqrt{2 \pi}} e^{-\frac{\phi^2}{2\sigma^2}}$. 
{This state can be written as a unitary operation on $|\psi_t\rangle$ ; i.e. $|\psi_t^\phi \rangle = e^{i \phi/2 (s^\dagger s - s^\dagger_\perp s_\perp)} |\psi_t \rangle$. The unitary $e^{i \phi/2 (s^\dagger s - s^\dagger_\perp s_\perp)}$ shifts the
azimutal angle of a qubit state of the form $(C_\alpha s^\dagger + e^{\phi_s}S_\alpha  s^\dagger_\perp)|\underline{0}\rangle$ by $\phi$.} When combining the unitary defining the setting choice and $e^{i \phi/2 (s^\dagger s - s^\dagger_\perp s_\perp)}$, we get the following expression for the emission modes as a function of the detected modes
\begin{equation}
	s = C_\alpha  A + S_\alpha  e^{i \phi_s+\phi} A_\perp
\end{equation}
\begin{equation}
	s_\perp = S_\alpha e^{-i \phi_s + \phi} A + C_\alpha  A_\perp
\end{equation}
and similarly for Bob.

\subsubsection{Summary}

The state which is effectively measured can be written as
\begin{equation}\label{eq:StateSummary}
	|\psi_{\alpha,\phi_s,\beta,\phi_a,\phi}\rangle = (1 - \tan(g)^2) e ^{(A^\dagger, A^\dagger_\perp) M 
	\left(\begin{array}{c} 
	B^\dagger \\
	B^\dagger_\perp \end{array} \right)}
	|\underline{0}\rangle
\end{equation}
with
\begin{equation}
	M = \tan(g) \left(\begin{array}{cc} 
	C_\alpha S_\beta e^{-i(\phi_a-\phi)}-S_\alpha e^{-i(\phi_s+\phi)}C_\beta & 
	-C_\alpha C_\beta-S_\alpha e^{i(\phi_s+\phi)}S_\beta e^{i(\phi_a-\phi)}\\
	S_\alpha e^{i(\phi_s+\phi)}S_\beta e^{-i(\phi_a-\phi)}+C_\alpha C_\beta  &
	-S_\alpha e^{i(\phi_s+\phi)}C_\beta+C_\alpha S_\beta e^{i(\phi_a-\phi)} \end{array} \right)
\end{equation}

It is measured according to a model where the POVM element associated to a click in detector A is given by
\begin{equation}
	\hat{D}_A(\eta_A) = 1-(1-p_{dc})(1-\eta_A)^{A^\dagger A}
\end{equation}
and similarly for $A_\perp$, $B$, and $B_\perp$. Given that $\phi$ is random and changes from run to run, we derive the probabilities of various measurement outcomes that we then average according to $p(\phi)$.

\subsection{Cross-Correlation Measurement}

We consider the cross-correlation measurement where Alice and Bob choose the settings $\alpha = \phi_s = 0 $ and $\beta = \phi_a = 0$ and measure
\begin{equation}
	g^{(2)}_{s,a} = \frac{\langle \psi_{\underline{0}} |\hat{D}_A(\eta_A) \hat{D}_B(\eta_B) |\psi_{\underline{0}}\rangle}
	{\langle \psi_{\underline{0}} | \hat{D}_A(\eta_A) |\psi_{\underline{0}}\rangle
	\langle \psi_{\underline{0}} | \hat{D}_B(\eta_B) |\psi_{\underline{0}}\rangle}
\end{equation}
where $|\psi_{\underline{0}\rangle} = |\psi_{0,0,0,0,\phi}\rangle$. A straightforward calculation (along the same lines as [53] 
 gives the following explicit expression
\begin{equation}
	g^{(2)}_{s,a} = \frac{
		1-(1-p_{dc})\frac{1-\mathrm{Th}_g^2}{1-\mathrm{Th}_g^2(1-\eta_A)}
		-(1-p_{dc})\frac{1-\mathrm{Th}_g^2}{1-\mathrm{Th}_g^2(1-\eta_B)}
		+(1-p_{dc})^2 \frac{1-\mathrm{Th}_g^2}{1-\mathrm{Th}_g^2(1-\eta_A)(1-\eta_B)}
		}
		{\left( 1-(1-p_{dc})\frac{1-\mathrm{Th}_g^2}{1-\mathrm{Th}_g^2(1-\eta_A)} \right)
		\left( 1-(1-p_{dc})\frac{1-\mathrm{Th}_g^2}{1-\mathrm{Th}_g^2(1-\eta_B)} \right)
		}
\end{equation}

From our experimental data, we can extract the following approximate values, as explained in Sec. \ref{sec:ExpParams}, $\mathrm{Th}_g^2 = 0.0022$ i.e. $g = 0.047$, $\eta_A = 0.1$, $\eta_B = 2.54 \times 10^{-4}$ (this includes the readout efficiency, i.e. anti-Stokes scattering probability knowing that a phonon was created) and $p_{dc} = 9 \times 10^{-6}$ (probability of dark counts per detection window), with which we get $g^{(2)}_{s,a} = 26.5$, in good agreement with the measured normalised coincidence. 
In this model, the decoherence of the collective molecular vibration within one mode at rate $\gamma_1$ manifests as an exponential decay of the vibration$\rightarrow$anti-Stokes conversion efficiency contained in $\eta_B$.

 

\subsection{The Interference Pattern for $\alpha = 0$}

\subsubsection{Twofold coincidence probability}

We consider the interference experiment in which the twofold coincidences on $A$ and $B$ are recorded when Alice fixes her measurement setting in the $s/s_\perp$ basis ($\alpha = \phi_s = 0$) while Bob rotates it in the x-z plane ($\phi_a = 0$). 
This situation corresponds to the blue curve of Fig.~3b in the main text.
We can find an explicit expression for these twofold coincidence probabilities
\begin{equation}
\begin{split}
	\langle \psi_{\underline{0},\beta}| \hat{D}_A(\eta_A) \hat{D}_B(\eta_B) |\psi_{\underline{0},\beta} \rangle 
	= \frac{1}{N}\left(
	1-(1-p_{dc})\frac{1-T_g^2}{1-T_g^2(1-\eta_A)}-(1-p_{dc})\frac{1-T_g^2}{1-T_g^2(1-\eta_B)} 
	\right. \\
	\left.  + (1-p_{dc})^2 \frac{2}{Ch_g^4} 
	\frac{1}{2-(2-\eta_A)(2-\eta_B)\mathrm{Th}_g^2 - \eta_A \eta_B C_{2\beta}\mathrm{Th}_g^2
	+2(1-\eta_A)(1-\eta_B)\mathrm{Th}_g^4}
	\right)
\end{split}
\end{equation}

The normalization coefficient $N$ accounts for the post-selection of events giving at least one click at each side, i.e.
\begin{equation}
\begin{split}
	N = 1-p(nc_A \& nc_{A_\perp} | \alpha = \pi / 4, \phi_s = 0, \phi)
	-p(nc_B \& nc_{B_\perp} | \beta , \phi_a = 0, \phi) \\
	+p(nc_A \& nc_{A_\perp}\& nc_B \& nc_{B_\perp} | \alpha = \pi / 4, \phi_s = 0,\phi_a = 0, \phi)
\end{split}
\end{equation}
where
\begin{equation}
	p(nc_A \& nc_{A_\perp} | \alpha, \phi_s, \phi) = (1-p_{dc})^2 
	\left( \frac{1-T_g^2}{1-T_g^2(1-\eta_A)} \right)^2
\end{equation}
\begin{equation}
	p(nc_B \& nc_{B_\perp} | \beta , \phi_a, \phi)= (1-p_{dc})^2 
	\left( \frac{1-T_g^2}{1-T_g^2(1-\eta_B)} \right)^2
\end{equation}
\begin{equation}
	p(nc_A \& nc_{A_\perp}\& nc_B \& nc_{B_\perp} | \alpha, \phi_s,\beta , \phi_a, \phi)
	= (1-p_{dc})^4 
	\left( \frac{1-T_g^2}{1-T_g^2(1-\eta_A)(1-\eta_B)} \right)^2
\end{equation}

and the notation $nc_A$ means ``no click in mode A", etc.

\subsubsection{Visibility of the Interference Pattern for $\alpha = 0$}

The visibility of the interference pattern is given by
\begin{equation}
	V_0 = \frac{\mathrm{max}_\beta \langle \psi_{\underline{0},\beta}| \hat{D}_A(\eta_A) \hat{D}_B(\eta_B) |\psi_{\underline{0},\beta} \rangle 
	- \mathrm{min}_\beta \langle \psi_{\underline{0},\beta}| \hat{D}_A(\eta_A) \hat{D}_B(\eta_B) |\psi_{\underline{0},\beta} \rangle	
	}
	{\mathrm{max}_\beta \langle \psi_{\underline{0},\beta}| \hat{D}_A(\eta_A) \hat{D}_B(\eta_B) |\psi_{\underline{0},\beta} \rangle 
	+ \mathrm{min}_\beta \langle \psi_{\underline{0},\beta}| \hat{D}_A(\eta_A) \hat{D}_B(\eta_B) |\psi_{\underline{0},\beta} \rangle	
	}
\end{equation}

Given the structure of (\ref{eq:StateSummary}), it is clear that $\langle \psi_{\underline{0},\beta}| \hat{D}_A(\eta_A) \hat{D}_B(\eta_B) |\psi_{\underline{0},\beta} \rangle $ is maximized for $\beta = 0$ and minimized for $\beta = \pi/2$. 
Using the same experimental parameters as above,  i.e. $\mathrm{Th}_g^2 = 0.0022$ i.e. $g = 0.047$, $\eta_A = 0.1$, $\eta_B = 2.54 \times 10^{-4}$ and $p_{dc} = 9 \times 10^{-6}$, we get $V_0 \approx 0.92$, in good agreement with the data of Fig.~3b (blue curve) in the main text.

\subsubsection{Visibility of the Interference Pattern for $\alpha = 0$ and Cross-Correlation Measurement}
Note that
\begin{equation}
\begin{split}
	|\psi_{\underline{0},\pi/2} \rangle = (1-\mathrm{Th}_g^2) e^{\mathrm{Th}_g^2 (A^\dagger B^\dagger e^{- i \phi/2} - A^\dagger_\perp B^\dagger_\perp e^{i \phi/2})} |\underline{0}\rangle = \\
	\underbrace{
	(1-\mathrm{Th}_g^2)^{1/2} e^{\mathrm{Th}_g^2 (A^\dagger B^\dagger e^{- i \phi/2}} |\underline{0}\rangle
	}_{|\psi_{AB}^{-\phi}\rangle}
	\otimes
	\underbrace{
	(1-\mathrm{Th}_g^2)^{1/2} e^{-\mathrm{Th}_g^2 (A^\dagger_\perp B^\dagger_\perp e^{i \phi/2}} |\underline{0}\rangle
	}_{|\bar{\psi}_{A_\perp B_\perp}^{+\phi}\rangle}
\end{split}
\end{equation}
implying
\begin{equation}
	\langle \psi_{\underline{0},\pi/2}| \hat{D}_A(\eta_A) \hat{D}_{B_\perp}(\eta_{B_\perp}) |\psi_{\underline{0},\pi/2} \rangle = \langle \psi_{A B}^{-\phi}| \hat{D}_A(\eta_A) |\psi_{A B}^{-\phi} \rangle
	\langle \bar{\psi}_{A_\perp B_\perp}^{+\phi}| \hat{D}_{B_\perp}(\eta_B) |\bar{\psi}_{A_\perp B_\perp}^{+\phi}\rangle
\end{equation}

Similarly, we have
\begin{equation}
	|\psi_{\underline{0}} \rangle = |\bar{\psi}_{A B_\perp}^{-\phi}  \rangle
	\otimes |\psi_{A_\perp B}^{+\phi}  \rangle
\end{equation}

We thus have
\begin{equation}
\begin{split}
	\langle \psi_{A B}^{-\phi}| \hat{D}_A(\eta_A) |\psi_{A B}^{-\phi} \rangle &=
	\mathrm{Tr}_{AB}(D_A(\eta_A)|\psi_{A B}^{-\phi} \rangle \langle \psi_{A B}^{-\phi} |) \\
	&=\mathrm{Tr}_{A}(D_A(\eta_A)\mathrm{Tr}_{B}(|\psi_{A B}^{-\phi} \rangle \langle \psi_{A B}^{-\phi} |))\\
	&=\mathrm{Tr}_{A}(D_A(\eta_A)\mathrm{Tr}_{B}(|\bar{\psi}_{A B\perp}^{-\phi} \rangle \langle \bar{\psi}_{A B\perp}^{-\phi} |)) \\
	&=\langle \bar{\psi}_{A B_\perp}^{-\phi}| \hat{D}_A(\eta_A) |\bar{\psi}_{A B_\perp}^{-\phi} \rangle \\
	&= \langle \psi_{\underline{0}}| \hat{D}_A(\eta_A) |\psi_{\underline{0}} \rangle
\end{split}
\end{equation}
and similarly
\begin{equation}
	\langle \bar{\psi}_{A B}^{+\phi}| \hat{D}_{B_\perp}(\eta_B) |\psi_{A B}^{-\phi} \rangle
	= \langle \psi_{\underline{0}}| \hat{D}_{B_\perp}(\eta_B) |\psi_{\underline{0}} \rangle
\end{equation}

From the previous equalities, we deduce
\begin{equation}
	\langle \psi_{\underline{0},\pi/2}| \hat{D}_A(\eta_A) \hat{D}_{B_\perp}(\eta_{B_\perp}) |\psi_{\underline{0},\pi/2} \rangle =
	\langle \psi_{\underline{0}}| \hat{D}_A(\eta_A) |\psi_{\underline{0}} \rangle
	\langle \psi_{\underline{0}}| \hat{D}_{B_\perp}(\eta_B) |\psi_{\underline{0}} \rangle
\end{equation}
and consequently
\begin{equation}
	V_{0}= \frac{g^{(2)}_{s,a} -1}{g^{(2)}_{s,a} +1}
\end{equation}

The previous formula holds for any efficiency. In particular, 
the temporal evolution of the visibility can be predicted from the evolution of $g^{(2)}_{s,a}$ and is thus ultimately limited by the decay of each single collective vibrational mode.

\subsection{Interference Pattern for $\alpha = \pi/4$}

We now consider the interference experiment in which the twofold coincidences on $A$ and $B_\perp$ are recorded when Alice fixes her setting to ($\alpha = \pi/4, \phi_s = 0$) while Bob rotates it in the x-z plane ($\phi_a = 0$), which corresponds to the red curve in Fig.~3b of the main text. 
This interference is sensitive to fluctuations in the phase of the superposition $\phi$ and thus allows to estimate its uncertainty. 
For fixed $\phi$, we have
\begin{equation}\label{eq:ExpVal_pi_4}
\begin{split}
	 \langle \psi_{\pi/4,\underline{0},\beta}| \hat{D}_A(\eta_A) \hat{D}_B(\eta_B) |\psi_{\pi/4,\underline{0},\beta} \rangle  
	 = \frac{1}{N}\left(
	1-(1-p_{dc})\frac{1-T_g^2}{1-T_g^2(1-\eta_A)}-(1-p_{dc})\frac{1-T_g^2}{1-T_g^2(1-\eta_B)} 
	\right. \\
	\left.  + (1-p_{dc})^2 \frac{2}{\mathrm{Ch}_g^4} 
	\frac{1}{2-(2-\eta_A)(2-\eta_B)\mathrm{Th}_g^2 - \eta_A \eta_B C_{2\phi}S_{2\beta}\mathrm{Th}_g^2
	+2(1-\eta_A)(1-\eta_B)\mathrm{Th}_g^4}
	\right)
\end{split}
\end{equation}
where the normalisation coefficient is given before. To take into account the uncertainty in $\phi$, we can first use a Taylor expansion of the term in Eq.~\ref{eq:ExpVal_pi_4}
\begin{equation}
	\int d\phi p(\phi)(1-p_{dc})^2 \frac{2}{\mathrm{Ch}_g^4} \times \frac{1}{\zeta - \xi (C_{2\phi-1}) }
	\approx (1-p_{dc})^2 \frac{2}{\mathrm{Ch}_g^4} \times \left(
	\frac{1}{\zeta} -\frac{2 \xi}{\zeta^2}\sigma^2 + O(\sigma^3) \right)
\end{equation}
where we introduced $\zeta = 2-(2-\eta_a)(2-\eta_B)\mathrm{Th)_g^2 - \eta_A \eta_B S_{2\beta}} \mathrm{Th}_g^2 + 2(1-\eta_A)(1-\eta_B)\mathrm{Th}_g^4$ and $\xi = \eta_a \eta_B S_{2\beta} \mathrm{Th}_g^2$. The visibility of the interference pattern is given by
\begin{equation}
	V_{\pi/4} = \frac{\mathrm{max}_\beta \langle \psi_{\pi/4, \underline{0},\beta}| \hat{D}_A(\eta_A) \hat{D}_{B\perp}(\eta_{B\perp}) |\psi_{\pi/4, \underline{0},\beta} \rangle 
	- \mathrm{min}_\beta \langle \psi_{\pi/4, \underline{0},\beta}| \hat{D}_A(\eta_A) \hat{D}_{B\perp}(\eta_{B\perp}) |\psi_{\pi/4, \underline{0},\beta} \rangle	
	}
	{\mathrm{max}_\beta \langle \psi_{\pi/4, \underline{0},\beta}| \hat{D}_A(\eta_A) \hat{D}_{B\perp}(\eta_{B\perp}) |\psi_{\pi/4, \underline{0},\beta} \rangle 
	+ \mathrm{min}_\beta \langle \psi_{\pi/4, \underline{0},\beta}| \hat{D}_A(\eta_A) \hat{D}_{B\perp}(\eta_{B\perp}) |\psi_{\pi/4, \underline{0},\beta} \rangle	
	}
\end{equation}

Given the structure of the state ~\ref{eq:StateSummary}, it is clear that $\langle \psi_{\pi/4, \underline{0},\beta}| \hat{D}_A(\eta_A) \hat{D}_{B\perp}(\eta_{B\perp}) |\psi_{\pi/4, \underline{0},\beta} \rangle $ is maximized for $\beta = \pi/4$ and minimized for $\beta = 3\pi/4$. 
Using again the same experimental parameter, i.e. $\mathrm{Th}_g^2 = 0.0022$ i.e. $g = 0.047$, $\eta_A = 0.1$, $\eta_B = 2.54 \times 10^{-4}$ and $p_{dc} = 9 \times 10^{-6}$, we can reproduce the visibility $V_{\pi/4} \approx 0.76$ obtained in Fig.~3b of the main text (red curve) for a width of the phase distribution $\sigma = 0.31$. 

\subsection{CHSH Value from the Interference Patterns}

Note that the correlation functions used to compute the CHSH value are given by
\begin{equation}
	E(\alpha,\beta) = p(+1 +1 |\alpha\beta)+ p(-1 -1 |\alpha\beta) - p(+1 -1 |\alpha\beta) - p(-1 +1 |\alpha\beta)
\end{equation}

The normalization implies $p(+1 -1 |\alpha\beta) + p(-1 +1 |\alpha\beta) = 1 - p(+1 +1 |\alpha\beta) - p(-1 -1 |\alpha\beta)$ and since $p(-1 -1 |\alpha\beta)= 1- p(+1|\alpha) - p(+1|\beta) + p(+1 +1 | \alpha\beta)$, we find
\begin{equation}
	E(\alpha,\beta) = 1-2p(+1|\alpha) -2p(+1|\beta) +4p(+1 +1 |\alpha\beta)
\end{equation}

When we record more than 2 clicks in one repetition of the experiment, we choose to bin the results according to the following rule: When detector $A$ clicks, Alice says that she gets $+1$ independently of the event on detector $A_\perp$. Similarly, when Bob gets a click on detector $B_\perp$, he says that he gets +1 independently of the event on detector B. This means that $p(+1 +1|\alpha\beta) = \langle \psi_{\alpha, \underline{0},\beta}| \hat{D}_A(\eta_A) \hat{D}_{B\perp}(\eta_{B\perp}) |\psi_{\alpha, \underline{0},\beta} \rangle $ while $p(+1 | \alpha) = \langle \psi_{\alpha, \underline{0},\beta}| \hat{D}_A(\eta_A) |\psi_{\alpha, \underline{0},\beta} \rangle$ and $p(+1 | \beta) = \langle \psi_{\alpha, \underline{0},\beta}| \hat{D}_{B\perp}(\eta_{B\perp}) |\psi_{\alpha, \underline{0},\beta} \rangle$. Given that we post-select the cases where at least one click is obtained at each side, we have
\begin{equation}
\begin{split}
	E(\alpha,\beta) = 1
	-\frac{2}{N_A} \langle \psi_{\alpha, \underline{0},\beta}| \hat{D}_A(\eta_A) |\psi_{\alpha, \underline{0},\beta} \rangle
	-\frac{2}{N_B} \langle \psi_{\alpha, \underline{0},\beta}| \hat{D}_{B\perp}(\eta_{B\perp}) |\psi_{\alpha, \underline{0},\beta} \rangle \\
	+\frac{4}{N} \langle \psi_{\alpha, \underline{0},\beta}| \hat{D}_A(\eta_A) \hat{D}_{B\perp}(\eta_{B\perp}) |\psi_{\alpha, \underline{0},\beta} \rangle 
\end{split}
\end{equation}
with $N_A = 1-p(nc_A\&nc_{A\perp}|\alpha,\phi_s,\phi)$ and $N_B = 1-p(nc_B\&nc_{B\perp}|\beta,\phi_a,\phi)$. Considering the angles maximizing the CHSH value for the singlet, we have
\begin{equation}
	\mathrm{CHSH} = E(0,\pi/8) + E(0,-\pi/8) + E(\pi/4,\pi/8) - E(\pi/4,-\pi/8)
\end{equation}
With the experimental parameters $\mathrm{Th}_g^2 = 0.0022$ i.e. $g = 0.047$, $\eta_A = 0.1$, $\eta_B = 2.54 \times 10^{-4}$ and $p_{dc} = 9 \times 10^{-6}$ and the phase uncertainty extracted above $\sigma = 0.31$, we find $\mathrm{CHSH} \approx 2.36$, in good agreement with the value measured close to zero delay (cf. Fig.~2 in the main text).

\subsection{Inferring Phonon Coherence Time from the CHSH Value} 

For the parameters of interest and in agreement with the measurement results, we checked that single photons are unpolarised on each side, meaning that the marginal probabilities of single photon detection are uniformly and randomly distributed onto the two detectors, whatever the measurement angle. This is shown in Fig.~3a of the main text. 
This means that the correlation functions only depends on the twofold coincidence probability

\begin{equation}
	E(\alpha,\beta) = \frac{4}{N}
	 \langle \psi_{\alpha, \underline{0},\beta}| \hat{D}_A(\eta_A) \hat{D}_{B\perp}(\eta_{B\perp})|\psi_{\alpha, \underline{0},\beta} \rangle -1
\end{equation}

Pure dephasing of the phononic qubit introduces a phase term $\bar{\phi}$ in the state similar to $\phi$ in Eq.~\ref{eq:Psi_t_phi}. This phase is different at each run and is distributed according to
\begin{equation}
	p(\bar{\phi}) = \frac{1}{\sqrt{2 \pi}} \bar{\sigma} e^{-\bar{\phi}^2 / 2 \bar{\sigma}^2}
\end{equation}
where the standard deviation $\bar{\sigma} = \sqrt{\gamma \Delta t}$ depends on both the dephasing rate $\gamma$ and the time duration $\Delta t$. From the previous analysis, we know that $\langle \psi_{0, \underline{0},\beta}| \hat{D}_A(\eta_A) \hat{D}_{B\perp}(\eta_{B\perp})|\psi_{0, \underline{0},\beta} \rangle$ is independent of $\bar{\phi}$ and hence $E(0,\beta)$ is independent of dephasing effects. From the previous perturbative approach, we also find that $\langle \psi_{\pi/4, \underline{0},\pm \pi/8}| \hat{D}_A(\eta_A) \hat{D}_{B\perp}(\eta_{B\perp})|\psi_{\\pi/4, \underline{0},\pm \pi/8} \rangle$ decays like $e^{-2 \bar{\sigma}^2}$. Given that $E(0,\pi/8) + E(0,-\pi/8)$ and $E(\pi/4,\pi/8) + E(\pi/4,-\pi/8)$ equally contribute to the CHSH value in the absence of phase noise, we have
\begin{equation}\label{eq:pure_dephasing_fit}
	CHSH(\Delta t) = \frac{CHSH}{2} (1 + e^{-2 \gamma \Delta t})
\end{equation}
The previous formula allows us to infer the coherence time $\gamma ^-1$ for the behavior of the CHSH parameter. Note that the prefactor CHSH is not constant in time since the detection efficiency of mode $B$ includes the phononic lifetime.

\section{Estimating experimental parameters}\label{sec:ExpParams}

\begin{itemize}
\item \textbf{Stokes detection efficiency $\eta_A$} The detection efficiency is the product of the avalanche photodiode efficiency, for which we use the value of 50\% from the manufacturer's test sheet, and the signal collection efficiency. Since our measurement is sensitive only to the spatial mode coupled into the single mode fiber used as a spatial filter, it is not necessary to consider the full spontaneous emission pattern in free space. Only the image of the single mode fiber into the sample is relevant to estimate the collection efficiency. This collection efficiency is therefore simply measured with the laser beam, which is well mode matched to the collection fiber, by measuring the power first just after the sample and second just before the detector. The additional loss due to internal reflection inside the diamond is estimated separately. This procedure yields $\eta_A \approx 0.1$ for the Stokes signal. 

\item \textbf{Squeezing parameter $g$} This detection efficiency, together with the measured count rate in detector $A$ of $18000$ counts$/$s, and the repetition rate of $80$~MHz, lets us estimate the average photon number $\langle n_A \rangle  = 2.25 \times 10^{-3}$ in the Stokes mode.  We then use the expression of $ \langle n_A \rangle $ to find the squeezing parameter $g = 0.047$.

\item \textbf{Dark count probability  $p_{dc}$} In the model, the dark count probability accounts not only for the intrinsic noise in the detectors (which has negligible impact on our measurements) but also, and foremost, for the anti-Stokes emission due to thermal phonons. Based on the anti-Stokes count rate at negative delays ($\sim 720 c/s$) and the detection time window we estimate the dark count rate: $p_{dc} = 9 \times 10^{-6}$.

\item \textbf{Vibration detection efficiency $\eta_B$} The detection efficiency of the vibrational mode $\eta_B$ accounts for both the probability of converting an existing vibration into anti-Stokes photon as well as the collection and detection losses for the anti-Stokes photon. We can extract it directly from our measurement by comparing the probability of detecting a Stokes photon, $P(A) = \langle n_A \rangle \eta_A $, to the probability of a coincidence $P(A \cap B) = \langle n_A \rangle \eta_A \eta_B $. With our coincidence rate of $4.58$ counts$/$s we obtain $\eta_B = \frac{P(A \cap B)}{P(A)} = 2.54 \times 10^{-4}$.
\end{itemize}



\section{Extracting the rate of pure dephasing}

\begin{figure}[h!]
	\centering
	\includegraphics[width=0.5\textwidth]{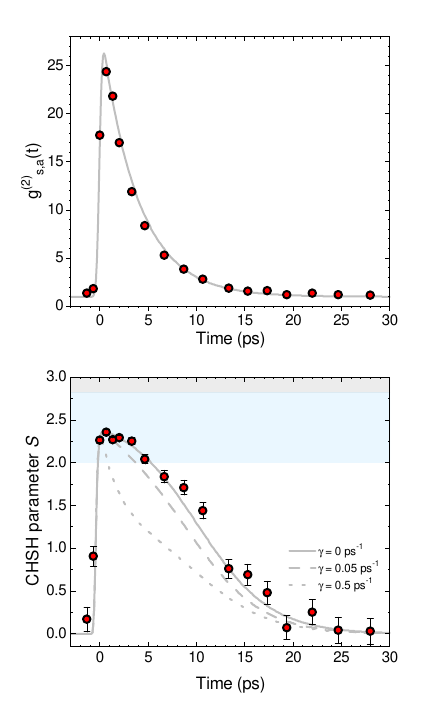}
	\caption{\label{fig:dephasing} \textbf{Comparison of the model with experimental data.} 
	\textbf{a}, Experimentally measured normalised Stokes--anti-Stokes cross-correlation $g^{(2)}_{s,a}(\Delta t)$ vs. the write--read time delay $\Delta t$, together with an exponential fit with a decay time constant $\tau=3.78$~ps, corresponding to the phonon lifetime. 
	\textbf{b}, Experimental CHSH parameter (as in main text, Fig.~2) overlaid with the curves computed from eq.~(\ref{eq:pure_dephasing_fit}) with the expression for $g^{(2)}_{s,a}(\Delta t)$ from panel \textbf{a}. Different values of the pure dephasing rate are shown, illustrating that our data are consistent with the decoherence of the vibrational qubit being dictated by population decay. The blue region, demarcated by $2 < |S| \leq 2\sqrt{2}$, certifies Bell correlations. 
	}
\end{figure}

We fit the measured $g^{(2)}_{s,a}(\Delta t)$ using a single exponential decay with time constant $\tau=3.78$~ps, convoluted with the instrument response function (Gaussian of width 200~fs). 

We then use eq.~(\ref{eq:pure_dephasing_fit}) to produce the expected curve for $S(\Delta t)$, using the fit of $g^{(2)}_{s,a}(\Delta t)$ for the temporal behaviour of $\eta_B$. 

The rate of pure dephasing is an adjustable, a priori unknown parameter. In Fig.~\ref{fig:dephasing}, we compare the measured CHSH parameter with the formula eq.~\ref{eq:pure_dephasing_fit} for various different dephasing rates ($\gamma$). The best agreement with the experimental data is found for $\gamma \ll \tau ^{-1}$, consistent with a lifetime-limited coherence time.


\section{Evolution of the CHSH parameter under ideal conditions}

We would like to address the following questions: if all technical noise could be eliminated from the photo-detection, including all background emission from the sample not related to vibrational Raman scattering, what would be the intrinsic dynamics how Bell parameter? For how long would Bell correlations persist? 

To answer these questions, we compute the temporal evolution of the Bell parameter using the theoretical model with idealized measurement, and with the experimentally determined vibrational energy decay rate, assuming the pure dephasing rate is much smaller and can be neglected. More explicitly, we use the following parameters:
\begin{itemize}
\item \textbf{Stokes detection efficiency $\eta_A$} We set the Stokes detection efficiency to unity.

\item \textbf{Vibration detection efficiency $\eta_B$} We set the initial value of the detection efficiency to unity, which then decays with the measured time constant corresponding to the phonon lifetime $\tau = 3.78$~ps.

\item \textbf{Dark count probability $p_{dc}$} We only include the anti-Stokes emission due to thermal phonons, which in the case of unit detection efficiency will be $p_{dc} = n_{th} = 1.7 \times 10^{-3}$.

\item \textbf{Squeezing parameter $g$} We find that the value that maximizes the time for which the CHSH inequality is violated is $g = 0.172$, corresponding to a mean photon number of $ \langle n_A \rangle = 0.030$
\end{itemize}

\begin{figure}[h!]
	\centering
	\includegraphics[width=0.5\textwidth]{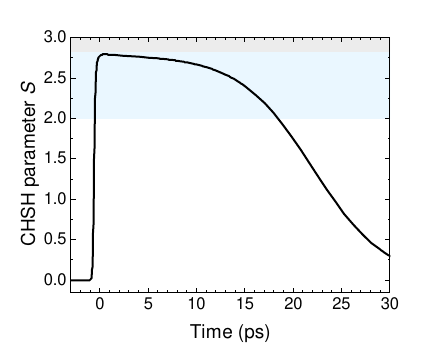}
	\caption{\label{fig:Ideal_CHSH} 
	\textbf{Evolution of the CHSH parameter under ideal experimental conditions.} The ideal conditions assume unit detection efficiency, noise  exclusively due to the thermal phonon occupancy, and optimal squeezing parameter.
	}
\end{figure}

In Fig.~\ref{fig:Ideal_CHSH} we show the time evolution of the CHSH parameter under ideal measurement. It is worth noting that the CHSH inequality is violated for $18.4$~ps, almost 5 times longer than the phonon lifetime, and more than twice the coherence time. 

This can be understood by noting that the Bell parameter at a given delay $\Delta t$ is ultimately limited by the value of $g^{(2)}_{s,a}(\Delta t)$, which is a measure of the signal-to-noise ratio for the conversion and detection of a heralded single phonon. 
As long as $\frac{g^{(2)}_{s,a}-1}{g^{(2)}_{s,a}+1} > \frac{1}{\sqrt{2}}$, or approximately $g^{(2)}_{s,a}(\Delta t) \geq 5.85$, the Bell parameter can exceed 2 (if all experimental imperfections reducing the two-photon interference visibility are mitigated). 
Therefore, even as $g^{(2)}_{s,a}(\Delta t)$ falls off exponentially with time, if its initial value is large enough (in the ideal case up to $1/n_\text{th}$) then  Bell correlation can be observed up to delays several times longer than the exponential coherence time. 

Note that this observation raises an interesting prospect. 
If the technical and background noises are significantly reduced, and the sample temperature  lowered, the initial value of $g^{(2)}_{s,a}$ can be made arbitrarily large. 
This would allow to put a more stringent bound on the pure dephasing rate $\gamma $, and maybe measure its magnitude even if it is much smaller than the exponential decay rate of the phonon. 
This a general comment that can be applied to other optomechanical systems in the quantum ground state as well. 

\section{Evaluation of the CHSH value from finite statistics}

\paragraph{CHSH as a game --}
In a CHSH test, Alice receives at each run a random bit $x=\{0,1\}$ and similarly for Bob $y=\{0,1\}.$ When Alice gets $x,$ she chooses the measurement setting $A_x$ while Bob chooses $B_y.$  For each setting choice, they receive a result $a=\{0,1\}$ for Alice and $b=\{0,1\}$ for Bob. They repeat the experiment many times so that they can evaluate 
\begin{equation}
\langle A_x B_y \rangle = p(a=b|A_xB_y) - p(a \neq b|A_xB_y). 
\end{equation}
The CHSH value is given by 
\begin{equation}
S=\langle A_0 B_0 \rangle + \langle A_0 B_1 \rangle + \langle A_1 B_0 \rangle - \langle A_1 B_1 \rangle.
\end{equation}
Such a test can be phrased as a game in which Alice and Bob receive $x$ and $y,$ respectively, as inputs and the winning condition is that their outputs satisfy $a\oplus b = x.y$ where $\oplus$ is the sum modulo 2. The winning probability $q$ relates to the CHSH value $S$ by 
\begin{equation}
q=\frac{4+S}{8}.
\end{equation}

\paragraph{Confidence interval on the mean value of winning probability--}
Let us see each experimental run as if a random variable $T_i$ was given. As an estimator of such a random variable $T_i,$ we choose 
\begin{equation}
T_i=\chi(a_i \oplus b_i = x_i.y_i)
\end{equation}  
with $\chi$ the indicator function, i.e. $\chi$(condition) = 1 if the condition is satisfied and 0 otherwise. Here $a_i$ is the result of Alice at run $i$ and similarly, $b_i, x_i$ and $y_i.$ Note that this estimator is unbiased. Indeed
\begin{eqnarray}
\nonumber
&\mathbf{E}(T_i)& = \sum_{a_i, b_i, x_i, y_i} T_i \, p(a_i, b_i, x_i, y_i)\\
\nonumber
&& = \sum_{a_i, b_i, x_i, y_i} T_i \, p(a_i, b_i, x_i, y_i)\\
\nonumber
&& = \sum_{a_i, b_i, x_i, y_i} T_i \, p(a_i, b_i|x_i, y_i) p(x_i, y_i)
\end{eqnarray}
and since $p(x_i, y_i)=1/4,$ 
\begin{eqnarray}
\nonumber
&\mathbf{E}(T_i)& = \frac{1}{4}\sum_{a_i, b_i, x_i, y_i} T_i \, p(a_i, b_i|x_i, y_i)\\
\nonumber
&& = \frac{1}{4} \big(p(a_i = b_i =0 |x_i \neq 1\, \text{and} \, y_i\neq 1) + p(a_i = b_i=1 | x_i \neq 1\, \text{and} \, y_i\neq 1 ) \\
\nonumber
&&  \, +  p(a_i = 0, b_i=1 |x_i =y_i=1)+ p(a_i = 1, b_i=0 |x_i = y_i=1)\big).
\end{eqnarray}
Note that
\begin{eqnarray}
\nonumber
&& p(a_i \oplus b_i =0 |x_i, y_i) = p(a_i = b_i =0 |x_i, y_i) + p(a_i = b_i=1 |x_i, y_i)=\frac{1}{2}(1+\langle A_{x_i} B_{y_i}\rangle )\\
\nonumber
&& p(a_i \oplus b_i =1 |x_i, y_i) = p(a_i = 0, b_i =1 |x_i, y_i) + p(a_i =1, b_i=0 |x_i, y_i)=\frac{1}{2}(1 - \langle A_{x_i} B_{y_i}\rangle )
\end{eqnarray}
Therefore 
\begin{equation}
\mathbf{E}(T_i) = \frac{1}{4} \left(\frac{4 + S_i}{2} \right) = q_i
\end{equation}
that is, the expectation of $T_i$ corresponds to the probability to win the game at run $i.$ We want to bound the average winning probability $\bar q = \frac{1}{n} \sum_{i} q_i.$ It was shown in Ref. [35] 
 that $[q_\text{min}, 1]$ is a confidence interval for $\bar q$ with
\begin{equation}
\label{hatq}
q_\text{min} = I_\alpha^{-1} (n \bar T, n(1-\bar T) +1) \, \text{with} \, \bar T = \frac{1}{n} \sum_i T_i
\end{equation}
where $0 \leq \alpha \leq 1/2 $ is the confidence level (e.g. $\alpha=0.01$ corresponds to a confidence level of $99\%$). Here we defined the inverse regularized incomplete Beta function $I^{-1},$ i.e. $I_y(a,b)=x$ for $y=I_x^{-1}(a,b).$ \\

Given a target confidence level $\alpha$,
the previous formula can be used to give a lower bound $S_\text{min}$ on the actual value of $\bar S$ using the following steps :\\
~~1 - Compute $T_i$ at each run using $T_i = \chi(a_i+b_i=x_i.y_i)$\\
~~2 - Deduce $\bar T = 1/n \sum_i T_i$\\
~~3 - Compute $q_\text{min}$ from the formula eq. \eqref{hatq} (for example with $\alpha=0.01$ for a confidence level of $99\%$)\\
~~4 - Deduce the lower bound $S_\text{min}$ on the mean CHSH value $\bar{S} = \frac{1}{n}\sum_i S_i$ using $S_\text{min} = 8*q_\text{min}-4.$ \\


\subsection{Example calculations}
We show in detail the calculation for $\Delta t = 0.66$~ps

We have $A_x = \{\alpha = 0,~ \alpha = \pi/2\}$ and $B_y = \{\varphi = -\pi/4 ,~ \pi/4\}$ as the settings for the experiment, and $a, b = 0$ corresponds to a click in the $+$ detector, while $ a, b = 1$ corresponds to either a click in the $-$ detector or the simultaneous clicking of both $+$ and $-$ detectors on one side (two-photon event).

The coincidence counts observed during the experiment for different settings are summarized in Table~\ref{tab:settings}.

\begin{table}[h!]
\begin{center}
\begin{tabular}{ |c|c|c|c|c|c|c|  } 
 \hline
 Setting & $n_{++}$ & $n_{+-}$ & $n_{-+}$ & $n_{--}$ & $n_{\pm +}$ & $n_{\pm -}$ \\ 
 \hline
 $\theta = 0, \varphi = -\pi/4$ & 1301 & 270 & 458 & 2034 & 0 & 0\\ 
 \hline
 $\theta = 0, \varphi = \pi/4$ & 1338 & 229 & 460 & 2006 & 1 & 0\\ 
 \hline
 $\theta = \pi/2, \varphi = -\pi/4$ & 388 & 1408 & 1549 & 694 & 0 & 1\\ 
 \hline
 $\theta = \pi/2, \varphi = \pi/4$ & 1468 & 494 & 328 & 1781 & 1 & 0\\ 
 \hline
\end{tabular}
\caption{\label{tab:settings} Coincidence counts}
\end{center}
\end{table}

Where $n_{\pm x}$ denotes the events involving two simultaneous coincidences in the Stokes measurement arm. There were no recorded events with simultaneous detections in the anti-Stokes arm.

We use this data to calculate $\bar T = 0.785$. We then compute $q_\text{min}$ for $\alpha = 0.01$ ($99\%$ confidence) and $\alpha = 5.733\times 10^{-7}$ ($5\sigma$ confidence) using \eqref{hatq}, and obtain $q_\text{min} = 0.788 $ and $q_\text{min} = 0.779$, respectively.

From this we conclude that the lower bound on $\bar{S}$ with $99\%$ confidence is $S_\text{min}=2.30$, and the lower bound with $5\sigma$ confidence is $S_\text{min}=2.23$, which comfortably violates the CHSH inequality.

\end{document}